\documentclass[a4paper,11pt]{article}
\pdfoutput=1 % if your are submitting a pdflatex (i.e. if you have
\usepackage{jheppub} % for details on the use of the package, please
\usepackage[T1]{fontenc} % if needed
\usepackage{hyperref}
\usepackage{ifpdf}
\usepackage{subfigure}
\usepackage{tikz}
\usepackage[compat=1.1.0]{tikz-feynman}
\usetikzlibrary{arrows,shapes}
\usetikzlibrary{trees}
\usetikzlibrary{matrix,arrows} 				% For commutative diagram
\usetikzlibrary{positioning}				% For "above of=" commands
\usetikzlibrary{calc,through}				% For coordinates
\usetikzlibrary{decorations.pathreplacing}  % For curly braces
\usepackage{pgffor}							% For repeating patterns

\usetikzlibrary{decorations.pathmorphing}	% For Feynman Diagrams
\usetikzlibrary{decorations.markings}
\tikzset{
	vector/.style={decorate, decoration={snake}, draw},
	fermion/.style={draw=black, postaction={decorate}}, 
	scalar/.style={dashed,draw=black, postaction={decorate}}}
\tikzstyle{block} = [draw, rectangle, 
minimum height=3em, minimum width=6em]

\usepackage{amssymb}
\usepackage{amsfonts}
\usepackage{epsf}
\usepackage{rotating}
\usepackage{graphicx}
\usepackage{amsmath}
\usepackage{fancyhdr}
\usepackage{lineno}
\usepackage{graphics}
\usepackage{pstricks}
\usepackage{color}
\usepackage{multirow}
\usepackage{array, booktabs, makecell}
\usepackage{siunitx, mhchem}

\newcommand{\nn}{\nonumber}
\newcommand{\lsim}{\mathrel{\mathop{\kern 0pt \rlap
			{\raise.2ex\hbox{$<$}}}
		\lower.9ex\hbox{\kern-.190em $\sim$}}}
\newcommand{\gsim}{\mathrel{\mathop{\kern 0pt \rlap
			{\raise.2ex\hbox{$>$}}}
		\lower.9ex\hbox{\kern-.190em $\sim$}}}

\newcommand{\be}{\begin{equation}}
\newcommand{\ee}{\end{equation}}
\newcommand{\bea}{\begin{eqnarray}}
\newcommand{\eea}{\end{eqnarray}}

\def\gev{\ensuremath{\mathrm{\,Ge\kern -0.1em V\,}}}
\def\tev{\ensuremath{\mathrm{\,Te\kern -0.1em V\,}}}

\title{\boldmath Right Handed Neutrinos, TeV Scale BSM Neutral Higgs and FIMP Dark Matter in EFT Framework}
\author[a]{Genevi\`{e}ve B\'{e}langer,}
\author[b]{Sarif Khan,}
\author[c,d]{Rojalin Padhan}
\author[c,d]{Manimala Mitra}
\author[c,d]{Sujay Shil}
\newcommand{\AddrHBNI}{
	Homi Bhabha National Institute, BARC Training School Complex, Anushakti Nagar, Mumbai 400094, India }
\affiliation[a]{Universite Grenoble Alpes, USMB, CNRS, LAPTh, F-74000 Annecy, France}
\affiliation[b]{Institut f\"{u}r Theoretische Physik,
Georg-August-Universit\"{a}t G\"{o}ttingen, Friedrich-Hund-Platz 1,
G\"{o}ttingen, D-37077 Germany}
\affiliation[c]{Institute of Physics, Sachivalaya Marg, Bhubaneswar, Pin-751005, Odisha}
\affiliation[d]{\AddrHBNI}
\emailAdd{belanger@lapth.cnrs.fr} 
\emailAdd{sarif.khan@uni-goettingen.de} 
\emailAdd{rojalin.p@iopb.res.in}
\emailAdd{manimala@iopb.res.in}
\emailAdd{sujayshil1@gmail.com}
\preprint{\textbf{IP/BBSR/2021-3}}
\abstract{We consider an effective field theory framework  with three standard
	model (SM) gauge singlet right handed neutrinos, and an additional SM gauge  singlet scalar field. The framework successfully generates eV masses of the light neutrinos via  seesaw mechanism, and  accommodates a feebly interacting massive particle (FIMP) as dark matter candidate. Two of the gauge singlet neutrinos   participate in  neutrino mass generation, while the third gauge singlet neutrino    is a FIMP dark matter. We explore the correlation between the {\it vev} of the gauge singlet scalar field which translates as mass of the  BSM Higgs,
	 and the mass of dark matter, which arises due to relic density constraint. We furthermore explore the constraints from the light neutrino masses
	 in this set-up. We chose the gauge singlet BSM Higgs in this framework  in the TeV scale. We  perform a detailed collider analysis to analyse  the discovery prospect of the TeV scale BSM Higgs through its di-fatjet signature, at a future $pp$ collider which can operate with $\sqrt{s}=100$ TeV c.m.energy.  }
\begin{document}
\maketitle
\flushbottom
\section{Introduction} \label{sec:I}
%%%%%%%%%%%%%%%%%%%%%%%%%%%%%%%%%%%%%%%%%%%%%%%%%%%%%%%%%%%%%%%%%%%%%%%%%
The Standard Model (SM) of particle physics, despite its  accurate predictions suffers from few serious deficits. Two of the most serious drawbacks emerge from the observation of light neutrino masses and their mixings, and the precise measurement of  dark matter (DM) relic abundance in the Universe. A number of
neutrino oscillation experiments have confirmed  that the solar and atmospheric neutrino
mass splittings are $\Delta m^2_{21} \sim10^{-5}\, \rm{eV}^2$,  $|\Delta m^2_{13}| \sim10^{-3}\, \rm{eV}^2$,  the Pontecorvo-Maki-Nakagawa-Sakata~(PMNS) mixing angles~\cite{Maki:1962mu,Pontecorvo:1967fh}
are $\theta_{12} \sim 33^\circ$, $\theta_{23} \sim 49^\circ$,  and $\theta_{13} \sim 8^\circ$ \cite{Esteban:2020cvm}. The light neutrinos being electromagnetic charge neutral can be Majorana particles. One of the profound mechanisms to generate Majorana masses of the light neutrinos is
seesaw, where tiny eV masses of the SM  neutrinos are generated from lepton number
violating (LNV) $d = 5$ operator \cite{Weinberg:1979sa,Wilczek:1979hc} through electroweak symmetry breaking.   Among the  different UV completed
theories that generate this operator,  type-I seesaw~\cite{Minkowski:1977sc,Mohapatra:1979ia,Yanagida:1979as,GellMann:1980vs} is possibly the most economic one, where   particle contents of the SM are extended to include gauge singlet  right handed neutrinos (RHNs). 
\paragraph{}Different models have been postulated, where the gauge singlet fermion state  can act as a DM  candidate~\cite{Kim:2006af,Kim:2008pp,LopezHonorez:2012kv}. A number of proposed models accommodate DM as a  weakly interacting massive particle (referred as WIMP), which is in thermal equilibrium with the rest of the plasma. However the null results from various  direct detection experiments cause serious tension for the WIMP paradigm, and therefore motivate to  explore  alternate DM hypothesis. One of such well-motivated mechanisms is freeze-in \cite{Hall:2009bx,McDonald:2001vt} production of DM. In this scenario, the DM has feeble interactions with the bath particles, and hence is referred as feebly interacting massive
particle (FIMP). The suppressed interaction  naturally explains the
non-observation of any direct detection signal. Moreover because of the very suppressed interaction, the FIMP  never attains
thermal equilibrium with the SM bath. In this scenario, DM is  produced from the decay and/or annihilation of the SM and Beyond Standard Model (BSM)
particles which are in thermal equilibrium \cite{Hall:2009bx}. 
FIMP DM has been explored in different contexts, see ~\cite{Biswas:2016bfo,Bandyopadhyay:2020ufc} for FIMP DM in $B-L$ model,  \cite{Elahi:2014fsa, Chen:2017kvz, Biswas:2019iqm, Bernal:2020bfj,Bhattacharya:2021edh} for EFT descriptions, \cite{Barman:2020plp} for discussion on all non-renormalizable operators upto dimension-8. Specific LHC signatures of FIMP DM have
been investigated in~\cite{Co:2015pka,Belanger:2018sti,Calibbi:2021fld,No:2019gvl} and others~\cite{Belanger:2020npe,Molinaro:2014lfa}.

\paragraph{}In this work we propose an effective field theory
set-up which includes a FIMP DM and explains  the origin of
light neutrino masses. The framework contains, in addition to SM
particles, three RHN states $N$ and one BSM scalar field $\chi$. Two of
the RHN states participate in the seesaw mechanism while the third RHN
is the FIMP DM. In our model, due to a discrete symmetry, the DM is 
completely stable. One of the specificity of our model is that the usual
renormalizable Dirac mass term for light neutrino mass generation  is absent and is only generated via an effective $d=5$ operator  $L\Phi N\chi/\Lambda$. Due to other sets of $d=5$ operator involving DM and scalars ($NN\chi\chi/\Lambda, NN\Phi\Phi/\Lambda$),  the DM is  mainly produced from the decay of scalars. Annihilation processes involving scalars or SM gauge bosons and fermions can also significantly contribute to DM production. The relative importance of decay and
annihilation processes for DM production strongly depends on the assumption on the reheating temperature of the early Universe.
We consider three different scenarios, {\it Scenario-I-III}, for the first two
only $d=5$ operators are responsible for both DM production in the early Universe, and generation of its mass, while in {\it Scenario-III} we add a bare mass term for the RHN DM and the other two RHN states. {\it Scenario-I} is a subset of {\it Scenario-II} where some of the operators are neglected for simplicity. 
\paragraph{}  We find that for the first two cases, a strong correlation exists between the \textit{vev} of $\chi$ and mass of DM, that emerges from the relic density constraint. While for the latter the correlation is somewhat relaxed. Demanding a TeV scale {\it vev} of $\chi$ and a   TeV scale heavy Higgs $H_2$ which offers a better discovery prospect of this model at collider,  a lighter KeV scale DM is in agreement with relic density constraint for {\it Scenario-I-II}.  For {\it Scenario-III} we find that a much heavier DM with  GeV scale mass is also consistent with a TeV scale {\it vev} of $\chi$, and in turn a TeV scale or lighter  BSM Higgs.  We furthermore study the impact of eV scale light neutrino mass constraint for these different scenarios. Using micrOMEGAs5.0~\cite{Belanger:2018ccd}, we perform a scan of all the relevant parameters such as, \textit{vev} of $\chi$, mass of DM,  reheating temperature, and show the variation of relic density. 

\paragraph{}Finally, we explore the collider signature of the BSM Higgs with TeV scale mass which actively participates in  DM production. For this, we consider a future $pp$ collider that can operate with c.m.energy $\sqrt{s}=100 $ TeV.  We consider  the decay of the BSM Higgs into two SM Higgs, followed by subsequent decays of the SM Higgs into $b \bar{b}$ states. For the TeV scale  BSM Higgs, the produced SM Higgs is highly boosted, thereby giving rise to collimated decay products. We therefore  study di-fatjet  final state as our model signature.  We consider a number of possible SM backgrounds including QCD, $WW/ZZ, W+j,Z+j, t\bar{t}$ which can mimic the signal. By judiciously applying selection cuts, we  evaluate the discovery prospect of the BSM Higgs. We find that a $3\sigma$ significance can be achieved for a 1.1 TeV BSM scalar with $30\, \text{ab}^{-1}$ luminosity for a large SM and BSM Higgs mixing angle.   
   
\paragraph{}The paper is organized as follows. In Section~\ref{model}, we describe  the  model and discuss  associated  DM phenomenology assuming  three different scenarios {\it Scenario I-III}, where DM is produced from the decay of the SM and BSM Higgs.  In Section~\ref{decvsannhi}, we discuss the contributions from both the decay and annihilation processes to the relic abundance and  show the variation of DM relic density w.r.t various parameters such as mass of DM, {\it vev} of the scalar field, and the reheating temperature. We perform  the collider analysis  of the BSM Higgs in di-fatjet channel  in Section~\ref{collider}. Finally, we conclude and summarize our findings in Section~\ref{conclu}.
%%%%%%%%%%%%%%%%%%%%%%%%%%%%%%%%%%%%%%%%%%%%%%%%%%%%%%%%%%%%%%%%%%%%%%%%%%%%%%%%%%%%%%%    
\section{The model \label{model}}
We consider an effective field theory framework with RHNs and one BSM scalar field, $\chi$, where we consider operators upto mass-dimension $d=5$. In addition to the SM particles,   the model  therefore contains three SM gauge singlet RHNs denoted as 
$N_{1,2,3}$, and one SM gauge singlet real scalar field $\chi$. The two RHNs $N_{1,2}$ generate eV Majorana masses of the SM neutrinos {\it via} seesaw mechanism, while  the state $N_3$ is a FIMP DM. The generic Yukawa Lagrangian with $N, \chi$ and the SM Higgs field $\Phi$    has the following form, 
\begin{eqnarray}
\mathcal{L}_{eff} & = & {M_{B}}_{ij}  N^T_i C^{-1} N_j+ \tilde{Y}_{ij} \bar{L_i} \tilde{\Phi} N_j + \tilde{Z}_{ij} N^T_i C^{-1} N_j  \chi+ \frac{c_{ij}}{\Lambda}  N^T_i C^{-1} N_j \chi^2+  \\ \nonumber   && \frac{c'_{ij}}{\Lambda}  N^T_i C^{-1} N_j \Phi^{\dagger} \Phi  +\frac{ Y_{ij}}{\Lambda} \bar{L}_i \tilde{\Phi} N_j \chi  +H.C,
\label{eq:caseC}
\end{eqnarray}
{where $\tilde{\Phi}=i\sigma_2 \Phi^\star$ and $M_B$ is the bare mass term of the RHNs. Other terms are the Yukawa interaction terms with couplings $\tilde{Y}_{ij}$, $\tilde{Z}_{ij}$, $c_{ij}$, $Y_{ij}$ and  $c^\prime_{ij}$, where $i,j=1,2,3$ are the generation indices. The parameter $\Lambda$ is the cut-off scale of this theory.}
In our subsequent discussions we do not consider  $\tilde{Y}, \tilde{Z}$ terms separately. These interaction terms can be obtained from $LN\Phi\chi$ and $NN\chi^2$ operators  via the \textit{vev} of $\chi$. A successful realization of the fermion state $N$ as a FIMP DM  demands  the coupling $\tilde{Z}, \tilde{Y}$ to be very tiny.  This can  naturally be obtained, if these terms are generated from  $NN\chi^2$ and $LN\Phi\chi$ operators, which feature  the $\frac{1}{\Lambda}$ suppression  factor.  Additionally, this is also to note that  by imposing a $Z_2$ symmetry under which  $\chi \to - \chi$,  $N_i \to - N_i$, and all other SM fields are invariant, the $\tilde{Y}, \tilde{Z}$ terms can be completely prohibited.  We impose  such a symmetry, hence our Lagrangian is 
		\begin{eqnarray}
		\mathcal{L}_{eff} & = & M_{B} N^T_i C^{-1} N_i + \frac{c_{ij}}{\Lambda}  N^T_i C^{-1} N_j \chi^2 + \\ \nonumber   && \frac{c'_{ij}}{\Lambda}  N^T_i C^{-1} N_j \Phi^{\dagger} \Phi  +\frac{ Y_{ij}}{\Lambda} \bar{L}_i \tilde{\Phi} N_j \chi  +H.C.
		\label{eq:genL}
		\end{eqnarray} 
	which only contains $d=5$ operators as interaction terms of RHNs. For simplicity we consider the Yukawa coupling matrices $c$,  $c^\prime$, and the bare mass matrix $M_B$ to be diagonal. As advertised before, among the $N_i$ states,  $N_3$ is DM. Therefore, the Yukawa matrix is required to have the following structure :
\begin{eqnarray}
Y= \begin{pmatrix} Y^{11}_{\nu} & Y^{12}_{\nu} & \epsilon \cr
Y^{21}_{\nu} & Y^{22}_{\nu} & \epsilon \cr
Y^{31}_{\nu} & Y^{32}_{\nu} & \epsilon \cr
\end{pmatrix}
\label{eq:YukawaA}
\end{eqnarray}
In the above, we consider  all $Y^{ij}_{\nu}$ ($i=1,2,3\,  \textrm{and}\,  j=1,2$) to be equal, while $\epsilon$ is required to satisfy the hierarchy $\epsilon \ll Y^{ij}_{\nu}$. The requirement of stability of DM over the age of the Universe forces the parameter $\epsilon$ to be orders of magnitude smaller than the other Yukawa couplings of the matrix $Y$. Note that the DM state $N_3$ can be made completely stable by imposing an additional $\mathcal{Z}_2$ symmetry, in which $N_3$ has  odd charge, and  all other fields are evenly charged. This  forbids the mixing between $N_3$ and light neutrino, {\it i.e.,}    $\epsilon = 0$. In this study we furthermore consider such a $\mathcal{Z}_2$ symmetry thereby making the DM state $N_3$  completely stable. The  interaction terms proportional to $\epsilon$ are hence absent in our case. 
%%%%%%%%%%%%%%%%%%%%%%%%%%%%%%%%%%%%%%%%%%%%%%%%%%%%%%%%%%%%%%%%%%%%%%%%%%%%%%%%%
\noindent\paragraph{\bf{Scalar Potential-}} As stated above,  the model also contains a gauge singlet scalar field $\chi$.  In addition to the Yukawa Lagrangian, the scalar field  $\chi$ also  interacts with the SM  Higgs doublet field $\Phi$ via the scalar potential, 
  \begin{eqnarray}
V(\chi, \Phi) &= & M_{\Phi}^2 \Phi^{\dagger}\Phi+ m^2_{\chi} \chi^2 + \lambda_1 (\Phi^{\dagger} \Phi)^2    + \lambda_2 \chi^4 + \lambda_3 
(\Phi^{\dagger} \Phi)\chi^2.
\label{eq:scalpot}
\end{eqnarray}
 The {\it $d=5$} terms $\frac{1}{\Lambda} {(\Phi^{\dagger} \Phi)}^2 \chi$,  and $\frac{1}{\Lambda}\chi^5$, as well as  $d=3$ term $\Phi^{\dagger} \Phi \chi$  are disallowed by the  above mentioned $Z_2$ symmetry. Therefore the scalar potential contains only renormalizable terms upto $d=5$.
  The spontaneous symmetry breaking (SSB) in this model is similar to the SM extension with an additional singlet scalar, which has been widely discussed in the literature \cite{Burgess:2000yq,Barger:2007im}. In order for the potential to be bounded from below, the couplings $\lambda_{1,2,3}$ should satisfy, 
\begin{align}
 4 \lambda_1 \lambda_2 - \lambda_3^2 > 0, \nonumber \\
 \lambda_{1,2} > 0.
\end{align}
We denote the $vev$s of $\Phi$ and $\chi$ by 
$v_{\Phi}$ and $v_{\chi}$, respectively. After  minimizing the potential $V(\chi,\Phi)$,  with respect to both the $vev$s, we obtain, 
\begin{eqnarray}
v^2_{\Phi} &= & \frac{4 \lambda_2 M_\Phi^2-2 \lambda_3 m^2_{\chi}  }{\lambda^2_3-4 \lambda_1 \lambda_2}, \\   v^2_{\chi} &= &\frac{4 \lambda_1 m_{\chi}^2-2 \lambda_3 M_\Phi^2}{\lambda^2_3-4 \lambda_1 \lambda_2}.
\label{eq:vevphichi}
\end{eqnarray}
{The $\lambda_3$-term in the potential enables mixing between $\chi$ and  $\Phi$ states. We denote the neutral Higgs component in the $\Phi$ multiplet as $H$.} The mass matrix between the two Higgs bosons in the basis 
$(H, \chi)$ is given by 
\begin{eqnarray}
{\cal M}(H, \chi) = 2 \left(\begin{array}{cc}
\lambda_1 v^2_{\Phi} & \lambda_3 v_{\Phi} v_{\chi}/2 \\ 
\lambda_3 v_{\Phi} v_{\chi}/2 & \lambda_2 {v^2_{\chi}}
\end{array}\right).
\label{eq:massmatrix}
\end{eqnarray}
The  mass eigenstates  $(H_1, H_2)$  are related to the $(H, \chi)$ states as
\begin{equation}
\left(\begin{array}{c}
H_1 \\
H_2 
\end{array}\right)= \left(\begin{array}{cc}
\cos \theta & -\sin \theta  \\
\sin \theta & \cos \theta
\end{array}\right)
\left(\begin{array}{c}
H \\
\chi 
\end{array}\right),
\label{mixingh}
\end{equation}
The mixing angle $\theta$  satisfies 
\begin{equation}
\tan 2\theta= \frac{ \lambda_3 v_{\chi} v_{\Phi}}{(\lambda_2 v^2_{\chi} -\lambda_1 v^2_{\Phi})}.
\label{theta}
\end{equation}
We denote the masses of the physical Higgs bosons as $M_{H_1}$ and $M_{H_2}$,
%\begin{widetext}
\begin{eqnarray}
M^2_{H_1}=\lambda_1 v^2_{\Phi}+ \lambda_2 v^2_{\chi}- \sqrt{(\lambda_1 v^2_{\Phi}- \lambda_2 v^2_{\chi})^2+\lambda^2_3 v^2_{\chi} v^2_{\Phi}},  \nonumber \\
M^2_{H_2}=\lambda_1 v^2_{\Phi}+ \lambda_2 v^2_{\chi}+ \sqrt{(\lambda_1 v^2_{\Phi}- \lambda_2 v^2_{\chi})^2+\lambda^2_3 v^2_{\chi} v^2_{\Phi}}.
\label{eq:mass} 
\end{eqnarray}
%\end{widetext}

Among the two Higgs states  $m^2_{H_1}< m^2_{H_2}$, \textit{i.e.}, $H_1$ acts as the 
lightest state. In our subsequent discussion,  we consider that  $H_1$ is SM-like Higgs with mass  $M_{H_1} \sim 125$ GeV. 
%The other Higgs state $H_2$  has a mass larger than $H_1$, satisfying  the above mentioned hierarchy. 
The interactions of $H_1$ and $H_2$ with the fermions and gauge bosons are given in the Appendix~(Section~\ref{App:AppendixA}). In this work, we consider that the BSM Higgs $H_2$ has a mass $M_{H_2} \sim $  TeV, or lower, and has a 
substantial mixing with the SM-like Higgs state $H_1$.  This large mixing facilitates the production of the BSM Higgs at colliders, which will
be discussed in Section~\ref{collider}.
%%%%%%%%%%%%%%%%%%%%%%%%%%%%%%%%%%%%%%%%%%%%%%%%%%%%%%%%%%%%%%%%%%%%%%%%%%%%%%%
\noindent\paragraph{\bf{FIMP Dark Matter}-}	As discussed above, we consider that the RHN state $N_3$ is a FIMP DM.  The state $N_3$, being gauge singlet only interacts via  Yukawa interactions $N_3 N_3 \chi^2/N_3 N_3\Phi \Phi$. Therefore, the production of 
	$N_3$  occurs primarily from the scalar states. In particular, the dominant contribution arises from the  decay of the BSM Higgs for a low reheating temperature $T_R<10^5$ GeV.  A number of annihilation channels, involving the SM/BSM Higgs and gauge boson also contribute to the relic density. For high reheating temperature,  the gauge boson annihilation channels   give dominant contributions, even larger than the decay contribution. The contributions from Higgs annihilation channels for a higher reheating temperature are also significantly large. In our discussion, we consider that the FIMP DM is lighter than the Higgs states $H_{1,2}$, such that, the decay of $H_{1,2}$ into 
	$N_3$ state is open.  The different channels that  lead  to the DM production  are 
	
\begin{itemize}
\item {Decay channels:  the Higgs decay $H_{1,2} \to N_3 N_3 $  generate the relic abundance. }
		
\item {Annihilation channels: the $2 \to 2$ annihilation channels, such as,  $WW/ ZZ \to N_3 N_3$, $ H_1 H_1 \to N_3 N_3$, $ H_2 H_2 \to N_3 N_3$, $H_1 H_2 \to N_3 N_3$   contribute to the production of $N_3$. We also consider annihilation of other SM particles such as, $b$ quark.}
		
\end{itemize}

In the subsequent discussion, we consider three different scenarios {\it Scenario I-III}, where we only consider the decay contribution of the SM and BSM Higgs. As stated above, this can be justified for a  lower reheating  temperature, for which the  annihilation processes give negligible contributions and DM production is primarily governed by the  decay of $H_1 \ \text{and} \  H_2 $. Among the three scenarios, in {\it Scenario-I} and {\it II}, we consider that the bare-mass terms of $N_{1,2,3}$  states are  zero. In this simplistic scenario the $d=5$ operator determines both the relic abundance, as well as DM mass, thereby leading to a tight correlation between mass of DM and \textit{vev} of $\chi$. In {\it Scenario-III}, we allow a non-zero bare mass term, that significantly alters the phenomenology.  We  analyse the constraints from DM relic density, and neutrino mass generation. We discuss the annihilation contributions in Section~\ref{decvsannhi}, where we depart from the assumption of a low reheating temperature.

% We also show the relative comparison between decay contribution and the annihilation contribution  to the relic density. We present  the variation of relic density with \textit{vev} of $\chi$, mass of DM as well as  reheating temperature $T_R$ and mass of the BSM Higgs $M_{H_2}$. For the discussion in Section~\ref{decvsannhi}, we use micrOMEGAs5.0~\cite{Belanger:2018ccd} for the evaluation of the relic density. 

%%%%%%%%%%%%%%%%%%%%%%%%%%%%%%%%%%%%%%%%%%%%%%%%%%%%%%%%%%%%%%%%%%%%%%%%%%%%%
\begin{table}[b]
	\centering
	\begin{tabular}{|l|l|l|l|l|l|l|}
		\hline
		&$M_{H_2}$  &$\sin\theta$  &$y$  & $c_{11}\ (c^\prime_{11})$ & $\qquad c_{33} \  (c^\prime_{33})$ & $M_{N_{1,2}}$      \\ \hline
		{\it Scenario-I}	&	$250$ GeV  & $0.1$ & $10^{-4}$ & $1 \ (0)$&\qquad $2.5 \times 10^{-6}\ (0)$ & $4 \times 10^{5}M_{N_3}  $   \\ \hline
		{\it Scenario-II}	&		$250$ GeV  & $0.1$ & $10^{-4}$ & $1 \ (1)$& $2.5 \times 10^{-6}\ (2.5 \times 10^{-6})$ & $4 \times 10^{5}M_{N_3}  $   \\ \hline
	\end{tabular}
	\caption{Parameters relevant for {\it Scenario-I}~(Fig.~\ref{case-AandB}) and {\it Scenario-II}~(Fig.~\ref{case-B}). } \label{tab:tab1}
\end{table}
%%%%%%%%%%%%%%%%%%%%%%%%%%%%%%%%%%%%%%%%%%%%%%%%%%%%%%%%%%%%%%%%%%%%%%%%%%%%%%%%%%%%%%%
\subsection{{\it Scenario-I}}
The RHN  states $N_i$  interact with the scalar field $\chi$, and the Higgs doublet $\Phi$ via the following Lagrangian:
%\begin{widetext}
\begin{equation}
 \mathcal{L}_{eff}= \frac{c_{ij}}{\Lambda}  N^T_i C^{-1} N_j \chi^2+ \frac{ Y_{ij}}{\Lambda} \bar{L}_i \tilde{\Phi} N_j \chi+\textrm{h.c.}
 \label{eq:eftA}
 \end{equation}
% \end{widetext}
In the above, $c_{ij}$ and $Y_{ij}$ are the Yukawa couplings, and $\Lambda$ is the cut-off scale of this theory.  As discussed in the previous section, we choose to work with a basis, in which the Yukawa coupling $c_{ij}$ is diagonal. The above Lagrangian, after electroweak symmetry breaking generates the following bi-linear terms involving  the light neutrinos, and RHNs $(\nu, N)$,
\begin{eqnarray}
\mathcal{L}_{eff} =  \frac{c_{ii}}{\Lambda}  N^T_i C^{-1} N_i v^2_{\chi}+  \frac{ Y_{ij}}{\Lambda} \bar{\nu}_i N_j v_{\Phi} v_{\chi}+ \textrm{h.c.}
\label{eq:eftmassA}
\end{eqnarray}
 The $\Lambda$ suppressed $d=5$ term $NN\chi\chi$ in Eq.~\ref{eq:eftA}  gives a natural explanation of the small interaction strength of the FIMP DM  $N_3$ with all other SM (and BSM) particles. As we consider the DM to be completely stable, therefore, only the $N_{1,2}$ states  participate in light neutrino mass generation. Below, we analyse  the contributions of $N_{1,2}$ in light neutrino mass, and the constraint from relic density.
%\begin{itemize}
%%%%%%%%%%%%%%%%%%%%%%%%%%%%%%%%%%%%%%%%%%%%%%%%%%%%%%%%%%
\noindent\paragraph{\bf{Neutrino Masses-}}
%The light neutrino mass will be generated by the seesaw mechanism
  The light neutrino masses will be generated due to the seesaw mechanism, where  two RHN states $N_{1,2}$ participate. In our case, $Y_{i3}=0$ and the Dirac mass matrix $M_D$  effectively reduces to a matrix of dimension $3\times2$. %\cro{For simplicity we consider $Y_{ij}=y$ for $j=1,2$.} 
  The Majorana mass matrix involving $N_{1,2}$ states is a $2\times2$ matrix.
 We denote the Dirac mass matrix by $M_D$ and the Majorana mass matrix of $N_1,\  N_2$ states by $M_R$, where 
 \begin{eqnarray}
 (M_D)_{\gamma \alpha}=\frac{ Y_{\gamma \alpha}}{\Lambda} v_{\Phi} v_{\chi},  ~~~ (M_R)_{\alpha \beta}= \frac{c_{\alpha \beta}}{\Lambda} v^2_{\chi} ~~~ (\alpha, \beta=1,2, \gamma=1,2,3).
 \label{eq:MDandmRsc1}
 \end{eqnarray}
 In the basis $\psi= (\nu_l, N_{R_1}^c, N_{R_2}^c)^T$, the  neutral lepton mass 
matrix becomes 
\begin{eqnarray}
M_{\nu} &=& \begin{pmatrix}
0 & M_D \cr
M^T_D & M_R
\label{matrix}
\end{pmatrix}.
\end{eqnarray}

The seesaw approximation $M_R > M_D$ translates into the hierarchy between the two {\it vevs} $v_{\chi} > v_{\Phi}$. The light neutrino and heavy Majorana mass matrix are given by, 
\begin{eqnarray}
m_{\nu} & = &  - M_D M^{-1}_R M^T_D,\, \,  M_N  \sim  M_R.
\label{eq:massneu}
\end{eqnarray}
These can further be re-written as, 
\begin{eqnarray}
m_{\nu}   \sim  - \frac{ v^2_{\Phi}}{\Lambda} { Y c^{-1}} Y^T, ~~~  M_{N}   \sim   \frac{c}{\Lambda} v^2_{\chi}.
\label{eq:massneuA}
\end{eqnarray}
In the above, $M_N=\text{diag}(M_{N_1}, M_{N_2})$ are the physical masses of the $N_{1,2}$ states, respectively. For simplicity we consider $M_{N_1}=M_{N_2}$ ({\it i.e.}, $c_{11}=c_{22}$) in all of the scenarios, {\it Scenario-I-III}. In terms of light neutrino masses, 
$M_{N_1}, M_{N_2}$ become  
\begin{eqnarray} 
M_{N_{1,2}} \sim  \frac{v^2_{\Phi} v^2_{\chi}}{\Lambda^2 } (YY^T m^{-1}_{\nu})\,  \textrm{GeV}.
\label{eq:caseAN12}
\end{eqnarray}
The active-sterile mixing matrix $V$ is related with Dirac mass matrix $M_D$ and Majorana mass matrix $M_N$ as
\begin{eqnarray} 
V \sim  M_D M_N^{-1} = \frac{v_{\Phi}}{ v_\chi} Y c^{-1}.
\end{eqnarray}
Next we discuss the relic abundance of $N_3$, where we consider the decays of $H_{1,2}$ as the primary production mode. 
%%%%%%%%%%%%%%%%%%%%%%%%%%%%%%%%%%%%%%%%%%%%%%%%%%%%%%%%%%%%%%%%%%%%%%%%%%%%%
\noindent\paragraph {\bf{Dark Matter Phenomenology-}}
In general, both decay and annihilation processes can produce $N_3$.
However, for the low reheating temperature that we consider in this
section, the production of $N_3$ occurs primarily from the decay of the scalar states. In our discussion, we consider that the DM is lighter than the Higgs states $H_{1,2}$, such that, the decay of $H_{1,2}$ into 
$N_3$ state is kinematically open. For illustrative purposes, in this section we consider the mass of the BSM Higgs $M_{H_2}=250 $ GeV. We have verified that  for larger TeV scale masses, such as $M_{H_2}=1.1$ TeV, the result presented in this section remains very  similar. A more extensive	investigation of the dependence of the relic density  on the mass of the	BSM Higgs is deferred to Section.~\ref{decvsannhi}.  The values of the BSM Higgs mass and mixing that we adopt to study  DM production  are consistent with the collider searches, which we will discuss in Section.~\ref{collider}.

\paragraph{} First note that the decay of the SM Higgs $H_1$ into $N_3$ state is non-negligible, only if the mixing between $H_{1,2}$ is sizeable. For the mixing $\theta \simeq 0$ the DM will be produced from the decay of $\chi \approx H_2$. From Eq.~\ref{eq:eftA} and using Eq.~\ref{mixingh}, the  interaction  Lagrangian of  $N_3$  with the SM and BSM Higgs reduces to,
%\begin{widetext}
\begin{eqnarray} 
\mathcal{L}_{N_3}  =   \frac{c_{33} v_{\chi}}{\Lambda}  \bar{N^c}_3  N_3 (-H_1 \sin \theta+ H_2 \cos \theta)+ \textrm{h.c.} 
 \label{eq:eftAN3}
 \end{eqnarray}

 {We define $\tilde{\lambda}_{1,2}$ as the couplings of $N_3$ with the $H_{1,2}$  states, respectively\footnote{We follow this notation for the rest of our discussion.}: 
\begin{eqnarray}
 \tilde{\lambda}_1= -\frac{c_{33} v_{\chi}}{\Lambda}  \sin \theta, ~~~  \tilde{\lambda}_2= \frac{c_{33} v_{\chi}}{\Lambda}  \cos \theta. 
 \label{eq:lambdas}
 \end{eqnarray}}
Since the bare mass term is zero, the mass of the DM state $N_3$ in this case is generated from the $c_{33} N_3N_3\chi\chi/\Lambda$ term once the $\chi$ state acquires \textit{vev} $v_{\chi}$. The mass of $N_3$ is therefore given by  
 \begin{eqnarray}
 M_{N_3}=\frac{v^2_{\chi} c_{33}}{\Lambda}
 \label{sce1dmmass}
 \end{eqnarray}
 and the couplings $\tilde{\lambda}_{1,2}$ can be expressed in terms of $M_{N_3}$,

\begin{eqnarray}
{\tilde{\lambda}_1 =}    -\frac{2 M_{N_3}}{ v_{\chi}}  \sin \theta, ~~~
 {\tilde{\lambda}_2 =} \frac{2 M_{N_3}}{ v_{\chi}}  \cos \theta
\label{eq:NNHcaseA}
\end{eqnarray}
Note that the interaction of $H_2$ with DM is governed by $\cos\theta$.
 The LHC Higgs signal strength measurements dictate $\sin \theta < 0.36$ i.e., $\cos \theta \simeq 1$~\cite{Sirunyan:2018koj}.  Therefore, for similar masses of  $H_{1,2}$, 
the BSM Higgs state $H_2$ primarily governs $N_3$ production due to a higher coupling strength $ {\tilde{\lambda}_2\gg  {\tilde{\lambda}_1}}$. However, for significantly heavier $H_2$, $H_1$ contribution in the DM relic density will be larger than from $H_2$. 
  
\begin{figure}[h]
 	\centering
 	\includegraphics[width=0.5\textwidth]{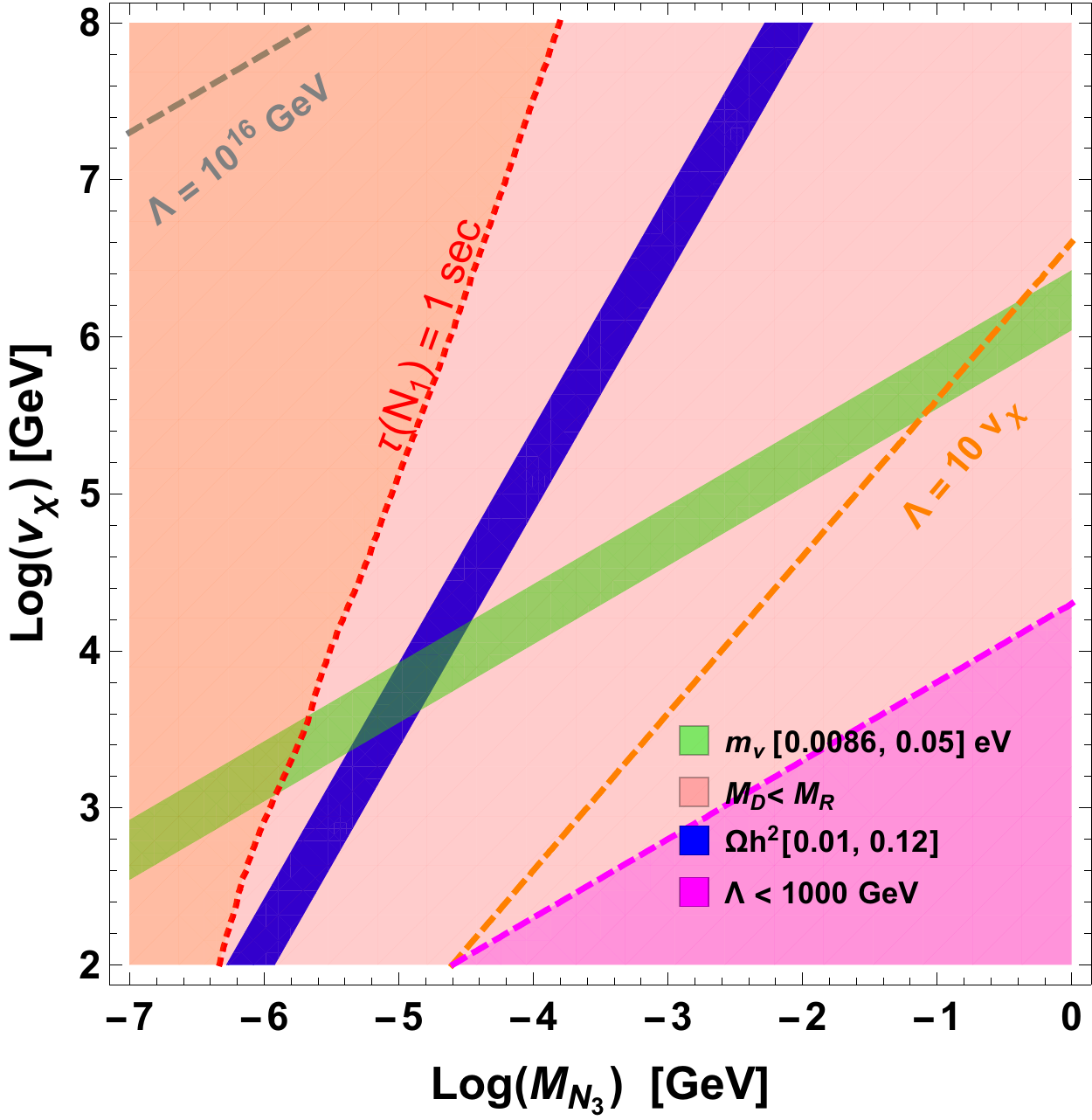}
 	\caption{This plot represents constraints on $M_{N_3}$ and $v_{\chi}$ for {\it Scenario-I}. The blue band corresponds to the variation of the relic density from decay in  between 0.01 and 0.12, where the latter satisfies  the experimental constraint~\cite{Aghanim:2018eyx}. The green region corresponds to light neutrino mass in between 0.0086 eV to 0.05 eV. The red line corresponds to lifetime of $N_{1,2}$ as  1 second. The orange line correspond to $\Lambda > 10.0 \, v_{\chi}$. Here we assume the mass of the BSM Higgs as $M_{H_2}=250$ GeV. }
 	\label{case-AandB}
 \end{figure}

As we are considering the decay contribution, the relic density  of the FIMP DM  can be expressed as \cite{Hall:2009bx}: 
\begin{eqnarray} 
\Omega_{N_3} h^2= \frac{2.18 \times 10^{27}}{g_s \sqrt{g_\rho}}  M_{N_3} \sum_{i=1}^{2} \frac{ g_{H_{i}} \Gamma_{H_{i}}}{M_{H_{i}}^2}.
\label{eq:darkmatterrelic}
\end{eqnarray}
In the above,  $g_{H_{i}}$ is the degrees of freedom~(d.o.f) of the decaying particle, $g_{s,\rho}\simeq103.857$ are the d.o.f of the Universe
related to entropy and matter. The 
partial decay widths $\Gamma_{H_{i}}$ for $H_{i} \to N_3 N_3$ are, 
\begin{eqnarray} 
\Gamma_{H_1} =  \frac{\tilde{\lambda}^2_1}{16 \pi} M_{H_1} ,  ~~~~ \Gamma_{H_2} =  \frac{\tilde{\lambda}^2_2}{16 \pi} M_{H_2}.   
\label{eq:partialW}
\end{eqnarray}
The measured relic abundance is $\Omega h^2 =0.1199\pm 0.0012$ at $68\%$ C.L~\cite{Aghanim:2018eyx}. 
Using the above equation and equating  Eq.~\ref{eq:darkmatterrelic} with the central value of the observed relic density, we obtain the constraints on the couplings as,
\begin{eqnarray} 
	\tilde{\lambda}_i= 1.66 \times 10^{-12} \sqrt{\frac{M_{H_i}}{M_{N_3}}}
	\label{eq:lambda},
	\end{eqnarray}
	where, we assume that the DM is entirely produced from either $H_{1,2}$.  {From Eq.~\ref{eq:NNHcaseA}, the couplings $\tilde{\lambda}_{1,2}$ depend on the mass of the DM, the {\it vev} $v_{\chi}$ of the BSM Higgs, and the mixing angle between two Higgs states $H_{1,2}$. Hence, using  Eq.~\ref{eq:NNHcaseA},~\ref{eq:partialW} in Eq.~\ref{eq:darkmatterrelic} and equating Eq.~\ref{eq:darkmatterrelic} with the observed relic density, we obtain  
	 a correlation between the {\it vev} of $\chi$, DM mass, and other physical parameters of this model, which are the Higgs mixing angle and mass of the Higgs.} 
	Taking into account both the $H_{1,2} \to N_3 N_3$ production modes, we find that, for a FIMP DM with  mass $M_{N_3}$, the required value of  $v_{\chi}$ has to satisfy the following constraint,
 \begin{eqnarray}
  v_{\chi} =  1.22\times 10^{12}  \ M_{N_3}^{3/2}  \left( \frac{\sin^2\theta}{ M_{H_1}} + \frac{\cos^2\theta}{ M_{H_2}} \right)^{1/2} \sim  10^{12}  \ M_{N_3}^{3/2}  \left( \frac{\theta^2}{ M_{H_1}} + \frac{1}{ M_{H_2}} \right)^{1/2},
  \label{eq:vchicaseAgeneral}
 \end{eqnarray}
where in the right-hand side we assume small values of $\theta$ and took
$\cos\theta \approx 1, \sin\theta \approx \theta$.
%In deriving the approximate expression, we use $\cos \theta \sim 1$.
 The strong correlation between $v_{\chi}$ and $M_{N_3}$ emerges, as both the DM mass and its production are governed by the same $N_3 N_3 \chi^2$ operator in the Lagrangian.
%\end{document}
In Fig.~\ref{case-AandB} we show this correlation. Before presenting the discussion on Fig.~\ref{case-AandB}, we note that, 
\begin{itemize}
\item

{The  contributions of $H_1 \to N_3 N_3$ and $H_2 \to N_3 N_3$ processes  to the relic density are: % the following forms 
\begin{eqnarray}
(\Omega_{N_3} h^2)_{H_1} \sim   \frac{\sin^2 \theta}{v^2_{\chi}}  \frac{M^3_{N_3}}{M_{H_{1}}},~~~  (\Omega_{N_3} h^2)_{H_2} \sim   \frac{\cos^2 \theta}{v^2_{\chi}}  \frac{M^3_{N_3}}{M_{H_{2}}}.
\label{eq:indomega}
\end{eqnarray}
Among these two, since the relic density from $H_2$ decay is proportional to $1/M_{H_2}$, therefore, for a very higher mass of the BSM Higgs state, its contribution can be sub-leading. On the other hand, for a much smaller value of $\sin \theta$, the contribution from $H_1$ decay can also be sub-leading.  }
\item
The ratio between the two contributions is 
\begin{eqnarray} 
\frac{(\Omega h^2)_{H_1}}{(\Omega h^2)_{H_2}}=  \frac{M_{H_2}}{M_{H_1}} \left(\frac{\sin^2\theta}{\cos^2\theta}\right) 
 \label{reltive_contri}
\end{eqnarray}

Therefore both the contributions can be comparable if the Higgs and BSM Higgs mixing angle $\theta$ satisfies $ \theta^2 \simeq 1.25 \times 10^2/M_{H_2}$.
Here we assume a small mixing angle, hence $\sin \theta \sim \theta$.

\end{itemize}
\paragraph{}In Fig.~\ref{case-AandB}, we show the constraint on  $v_{\chi}$ and on the DM mass $M_{N_3}$  that arise from requiring the relic density to lie in the range $0.01<\Omega h^2<0.12$ (blue region), namely  we allow for $N_3$ to account for  $10-100\%$
of the DM abundance.
Here we include both $H_1 \to N_3 N_3$ and $H_2 \to N_3 N_3$ contributions to the relic density even though the $H_2$ contribution dominates as we fixed $M_{H_2}=250$ GeV.
 For this figure, we use the benchmark parameters given in Table.~\ref{tab:tab1}. Additionally, we also show the constraint from eV light neutrino mass in the same plot. For simplicity here and in other figures as well, we consider the light neutrino mass matrix $m_{\nu}$  (also $M_D$ and $M_R$) as a parameter, and impose neutrino mass constraint. Hence,  $y$ denotes the Dirac Yukawa coupling  parameter in Table.~\ref{tab:tab1}.
 The green shaded region is compatible
 	with eV light neutrino masses \footnote{Two of the RHN states $N_{1,2}$ will participate in neutrino mass generation. Hence, the lightest neutrino mass $m_{1}/m_{3}=0$ depending on normal/inverted mass hierarchy in the light neutrino sector.   We therefore  vary $m^2_\nu$ in between solar and atmospheric mass square splittings, where we consider  ${\Delta m^2_{21}} = {7.42\times10^{-5}\,} \rm{eV}$ and  ${|\Delta m^2_{13}|} = {2.517\times10^{-3}\,} \rm{eV}$~\cite{Esteban:2020cvm}.}, $0.0086  \ \text{eV}< m_\nu< 0.05 \  \text{eV}$, while the
 	seesaw approximation $M_R > M_D$ is satisfied in the entire plot.
 	 We further note that, following Eq.~\ref{sce1dmmass},  for higher $v_{\chi}$ and lower  $M_{N_3}$ the cutoff scale $\Lambda$ increases. The brown dashed line in the  top left corner denotes $\Lambda=10^{16}$ GeV. We also show the line corresponding to $\Lambda =10v_{\chi}$ by orange dashed line. The magenta dashed line, assuming $\Lambda>1$ TeV rules out the region with large $M_{N_3}$ and low $v_\chi$ (magenta shaded region).
 	 
	 \paragraph{}The constraint from the relic density  depends on the Yukawa coupling $c_{33}$, which has been rewritten in terms of  $M_{N_3}$. However, the constraint from eV light neutrino masses depend on other parameters, such as, $M_{N_{1,2}}$, and hence the couplings $c_{11}, c_{22}$ as well as the Dirac Yukawa $y$.   As can be seen from the figure, to satisfy the observed DM relic  density, the required value $v_{\chi}$ increases with DM mass $M_{N_3}$. For GeV scale $M_{N_3}$, one needs $v_{\chi}>10^8$ GeV. This naturally leads to a very heavy BSM Higgs with mass\footnote{ $M_{H_2}$ can not be much larger than $v_{\chi}$ due to perturbitivity bound of $\lambda_2$.}   $M_{H_2}\sim v_\chi>10^8$ GeV for the quartic scalar coupling $\lambda_2\sim1$. This very heavy BSM Higgs does not have any detection prospect at collider. Contrary to that, the coupling $\lambda_2$ needs to be extremely tiny 
$\lambda_2<10^{-12}$ to accommodate $M_{H_2} \sim \mathcal{O}(100)$ GeV, which has better discovery prospect at the ongoing and future colliders. This unnatural fine-tuning relaxes, if  the DM mass is $M_{N_3} \sim$  KeV.  As  can be seen from the figure, DM  $N_3$ with few KeV mass is consistent with a $v_{\chi} \sim $ TeV and $\lambda_2 \sim \mathcal{O}(0.1)$ and hence $M_{H_2} \sim \mathcal{O}(100)$ GeV. 
In conclusion, in {\it Scenario-I} we find that  relic density
constraint prefers a KeV scale DM and a TeV scale $v_\chi$ to naturally accommodate  a BSM Higgs at the TeV scale or below. 
%We  conclude this discussion with the observation that a KeV scale DM in {\it Scenario-I}, along with a TeV scale or lighter BSM Higgs  is a more natural choice, as compared to the heavier DM along with a TeV scale/lighter BSM Higgs. 
%This is also clearly visible from the figure, that  for the above mentioned parameters it is possible to satisfy both the light neutrino mass constraint and observed relic density.  
%%%%%%%%%%%%%%%%%%%%%%%%%%%%%%%%%%%%%%%%%%%%%%%%%%%%%%%%%%%%%%%
\noindent\paragraph {\bf{$N_1$ and $N_2$  lifetime-}}
%%%%%%%%%%%%%%%%%%%%%%%%%%%%%%%%%%%%%%%%%%%%%%%%%%%%%%%%%%%%%%%%%%%%%
Before concluding the section we also discuss the lifetime of $N_{1,2}$. For the range of the relevant parameters that we consider in Fig.~\ref{case-AandB}, 
the mass of the RHN states $N_{1,2}$ vary from $10^{-2}-10^5$ GeV, while the mixing $V$ ranges from $V \sim 10^{-4} -10^{-10}$. Here for simplicity, we  consider $V$ as a parameter. For large mixing, $N_{1,2}$ state will thermalise and their decays would be constrained from the Big Bang Nucleosynthesis~(BBN). While a detailed evaluation of the BBN bound is beyond the scope of this present paper, we however show the lifetime contour in Fig.~\ref{case-AandB} that corresponds to $\tau(N_{1,2}) \sim 1$ sec. The  two RHN states $N_{1,2}$ decay to various final states via their mixing $V$ with the active neutrinos.  For masses much smaller than the pion mass, the decay mode would be $\nu \gamma$ and $\nu \nu \nu$. The decay width and lifetime for  this mass range are
 \begin{eqnarray}
 \tau^{-1}_{N_{1, 2}} &=& \Gamma_{N_{1, 2}} \simeq \frac{G^2_{F} M^5_{N_{1,2}}}{96 \pi^3} V^2 \nn \\
 &=& 5.16 \times 10^{-24} \left(\frac{M_{N_{1,2}}}{keV}\right)^5\,
 \left( \frac{V^2}{10^{-7}} \right) s^{-1}. \label{tauN1}
 \end{eqnarray}
For larger mass range $M_{N_{1,2}}> m_{\pi}+m_e$, additional decay modes $N \to l \pi^{\pm}$, $N \to l B^{\pm}/K^{\pm}$, and others will be open.  For even higher mass range 
$M_{N_{1,2}} > M_W, M_Z, M_{H_{1,2}}$, the two body modes $N \to l W, \nu Z, \nu H_{1,2}$ will be open.  The expressions for these decay widths are:

\begin{eqnarray}
\Gamma_{N_{1, 2}} (N \to l \pi) \simeq \frac{G^2_{F} M^3_{N_{1,2}}}{96 \pi^3} V^2 \nn\\
 \label{twobodylpi}
\end{eqnarray}
\begin{eqnarray}
\Gamma_{N_{1, 2}} (N \to l W) \simeq \frac{g^2_{F} M^3_{N_{1,2}}}{64 \pi M^2_W} V^2 \left(1-\frac{M^2_W}{M^2_N}\right)^2\left(1+2\frac{M^2_W}{M^2_N} \right) 
 \label{twobodyllW}
\end{eqnarray}
\begin{eqnarray}
\Gamma_{N_{1, 2}} (N \to \nu Z) \simeq \frac{g^2_{F} M^3_{N_{1,2}}}{128 \pi M^2_W} V^2 \left(1-\frac{M^2_Z}{M^2_N}\right)^2\left(1+2\frac{M^2_Z}{M^2_N} \right) 
 \label{twobodyllW}
\end{eqnarray}
\begin{eqnarray}
\Gamma_{N_{1, 2}} (N \to \nu H_{1,2}) \simeq \frac{g^2_{F} M^3_{N_{1,2}}}{128\pi M^2_W} V^2 \left(1-\frac{M^2_{H_{1,2}}}{M^2_N}\right)^2 
 \label{twobodyllW}
\end{eqnarray}
\paragraph{}  We evaluate the lifetime of $N_{1,2}$ assuming $M_{N_1}=M_{N_2}$ and show the contour of $\tau(N_{1,2})=1$ sec in Fig.~\ref{case-AandB} by the red line. Part of the region in the left side of the red line can be constrained  from BBN as $N_{1,2}$ thermalise, and  the decay of $N_{1,2}$ happens after $\tau(N_{1,2})=1$ sec. We estimate that for the region of Fig.~\ref{case-AandB} in agreement with both relic
  	density and light neutrino mass, for which $M_{N_3}\sim 10-30$ KeV and  $v_{\chi} \sim 4-10$ TeV, the cut-off scale $ \Lambda \sim 10^{7}$ GeV, and the mixing angle $V \sim 10^{-6}$. Thus, $\tau(N_{1,2})<1$ sec, see Eq.~\ref{tauN1}, and the decay of $N_{1,2}$ in the early Universe occurs before BBN.
 
 % We estimate  that for the  overlapping mass range of the FIMP DM in between 1-10 KeV and  $v_{\chi} \sim 1-10$ TeV shown in Fig.~\ref{case-AandB}, which is  in agreement with both DM and neutrino mass constraint,  the cut-off scale $ \Lambda \sim  10^{7}$ GeV, and mixing angle $V \sim 10^{-5}-10^{-6}$. From Eq.~\ref{tauN1}  the lifetime of $N_{1,2}<1$ sec. \crd{Hence  the decay of $N_{1,2}$ in the early Universe  occurs before BBN. }
%\end{itemize}
%%%%%%%%%%%%%%%%%%%%%%%%%%%%%%%%%%%%%%%%%%%%%%%%%%%%%%%%%%%%%%%%%%%%%%%%%%%%%%%
\subsection{\it Scenario-II}  
{We consider that in addition to the $ N^T C^{-1}  N \chi^2$ term,  the Yukawa  Lagrangian contains the term $ N^T C^{-1}  N \Phi^{\dagger} \Phi$. This is a more generic choice, as {\it Scenario-I} can be realised as only a special case of {\it Scenario-II} with $c^{\prime}=0$. } The Lagrangian has the following terms:
  %\begin{widetext}
 \begin{eqnarray}
 \mathcal{L}_{eff} = \frac{c_{ij}}{\Lambda}  N^T_i C^{-1} N_j \chi^2+ \frac{c^{\prime}_{ij}}{\Lambda}  N^T_i C^{-1} N_j \Phi^{\dagger} \Phi+ 
&& \frac{Y_{ij}}{\Lambda} \bar{L}_i\tilde{\Phi} N_j  \chi+ \textrm{h.c.}
 \label{eq:eftB}
 \end{eqnarray}
 In this scenario, the RHN neutrino masses get contributions from both the $NN\chi^2$  and $NN\Phi^{\dagger} \Phi$ terms. As before, we consider $c, c'$ to be diagonal matrix. %In the $c^{\prime} \to 0$ limit, one obtains the scenario presented in {\it Scenario-I}. 
 The mass matrix of the two  RHN's is 
  \begin{eqnarray} 
 (M_R)_{\alpha \beta}= \frac{c_{\alpha \beta}}{\Lambda} v^2_{\chi}+\frac{c^{\prime}_{\alpha \beta}}{\Lambda} v^2_{\Phi} ~~~ (\alpha, \beta=1,2)
 \label{eq:caseBmass}
 \end{eqnarray}
 The DM has a mass 
   \begin{eqnarray} 
 M_{N_3}= \frac{c_{33}}{\Lambda} v^2_{\chi}+\frac{c^{\prime}_{33}}{\Lambda} v^2_{\Phi}
 \label{eq:caseDMmass}
 \end{eqnarray}
 The Dirac mass matrix has the same expression as in the previous
 section, Eq.~\ref{eq:MDandmRsc1}, and the physical mass matrix of $N_{1,2}$  follows $M_N \sim M_R$.  With the  seesaw condition $M_R > M_D$,  the light neutrino mass matrix  has a similar expression as Eq.~\ref{eq:massneu}. Below, we consider  the couplings $c \sim c^{\prime} \simeq c_0 $. Therefore, the light neutrino mass matrix receives a correction of $\mathcal{O}(\frac{v^2_{\Phi}}{v^2_{\chi}})$. The light neutrino and heavy RHN mass matrix have the following form, 
  \begin{eqnarray}
 m_{\nu} &\sim&  - {Y}\frac{1}{ c_0 \Lambda} v^2_{\Phi} Y^T (1- \frac{v^2_{\Phi}}{v^2_{\chi}}), \nonumber  \\
   M_N &\sim&  \frac{c_0}{\Lambda} v^2_{\chi} (1+\frac{v^2_{\Phi}}{v^2_{\chi}})
\label{eq:massneuB}
\end{eqnarray}

Similar to the previous scenario, the RHN $N_3$ in this case is the FIMP DM. The  particle is primarily produced from the two Higgs states $H_{1,2}$. 
 The  couplings of $N_3$ with $H_{1,2}$ states have the following form:
  \begin{eqnarray}
 \tilde{\lambda}_1 :    -\frac{2 v_{\chi} c_{33}}{\Lambda}  \sin \theta + \frac{2 v_{\Phi} c^{\prime}_{33}}{\Lambda}  \cos \theta 
 \label{eq:DMHiggsinteractionB1}
 \end{eqnarray}
 \begin{eqnarray}
 \tilde{\lambda}_2:  \frac{ 2 v_{\chi} c_{33}}{\Lambda} \cos \theta+ \frac{  2 v_{\Phi} c^{\prime}_{33}}{\Lambda}   \sin \theta. 
 \label{eq:DMHiggsinteractionB2}
 \end{eqnarray}

We first discuss  two extreme scenarios, 
\begin{itemize}
\item the $N_3 N_3 H_1$ coupling $\tilde{\lambda}_1$ is zero, i.e., the FIMP is produced only from $H_2$ decay.
\item the $N_3 N_3 H_2$ coupling $\tilde{\lambda}_2$ is zero, i.e., the FIMP is produced from $H_1$ decay. \end{itemize}

 For the subsequent discussions, we consider the couplings $c$ and $c^{\prime}$ independently. \\

 (a) In the first scenario, the DM is entirely produced from the Higgs state $H_2$. Imposing $\tilde{\lambda_1}=0$ in Eq.~\ref{eq:DMHiggsinteractionB1} leads to
\begin{eqnarray}
\tan \theta = \frac{v_{\Phi} c^{\prime}_{33}}{v_{\chi} c_{33}}. 
\label{eq:caseBconstraint}
\end{eqnarray}
 Using Eq.~\ref{eq:caseBconstraint}, $ \tilde{\lambda}_2$ can be simplified as, 
	\begin{eqnarray}
	  \tilde{\lambda}_2=  \frac{ 2 M_{N_3}}{(v_{\chi} \cos \theta + v_{\Phi} \sin \theta)}.
	\label{eq:NNHcaseB1}
	\end{eqnarray}
	Using  the above coupling in  Eq.~\ref{eq:lambda} we obtain the constraint on the {\it vev} $v_{\chi}$ from $\Omega h^2=0.12$,
	 \begin{eqnarray}
	v_{\chi}=\frac{1.22\times 10^{12}  \ M_{N_3}^{3/2} }{ M_{H_2}^{1/2} \cos\theta} - \frac{ v_{\Phi}\sin\theta}{\cos\theta}.
	\label{eq:vchicaseB1}
	\end{eqnarray}

\begin{figure*}[t]
	%\centering
	\includegraphics[width=0.55\textwidth]{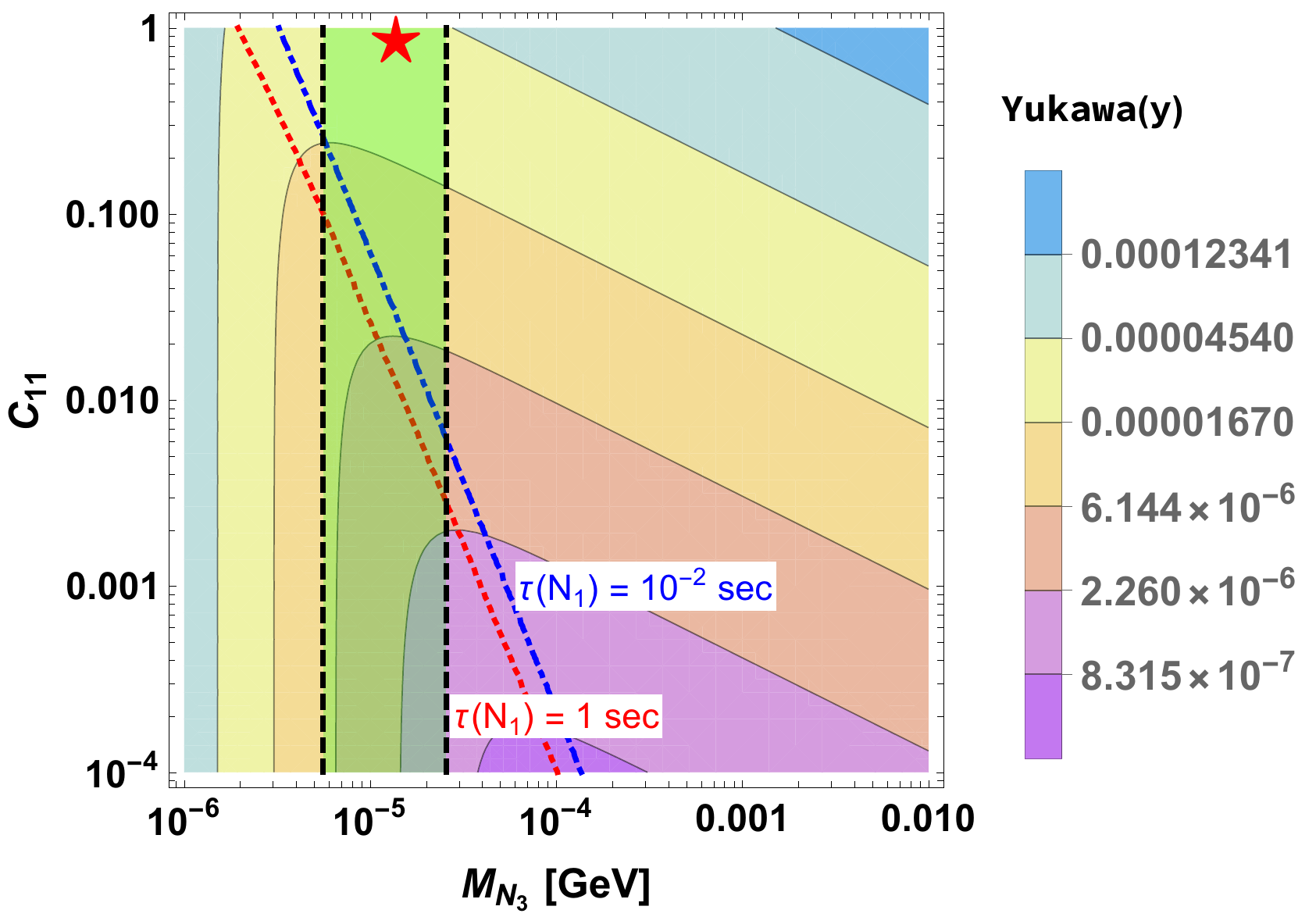}
	\includegraphics[width=0.55\textwidth]{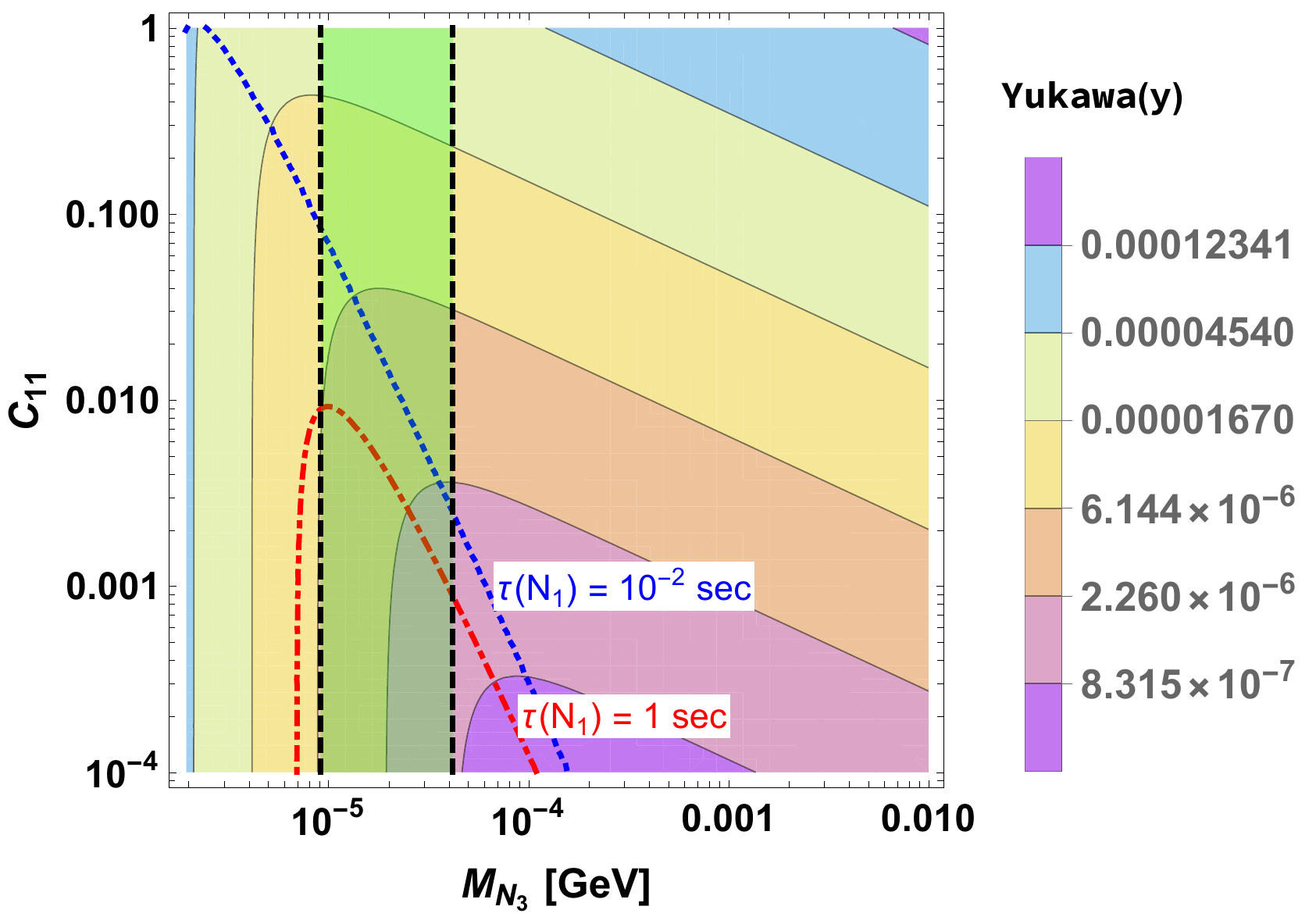}
	\caption{{Left and Right panel:  the figures correspond to  {\it scenario-IIa}  for  two different Higgs masses $M_{H_2}=$250 GeV and 1.1 TeV, respectively. The color bar indicates the variation of Dirac Yukawa coupling $y$ w.r.t  the variation of mass of FIMP DM $ M_{N_3}$ and the Yukawa coupling $c_{11}$. The green band indicates  variation of $v_{\chi}$  in between 1 TeV to 10 TeV (from left to right).  See the texts for additional details.}} 
	\label{fig:case-B1}
\end{figure*}

	Written in this way, the relic density constraint does not directly depend on the interaction coupling $c_{33}$ of the Yukawa Lagrangian, the dependency is only via $M_{N_3}$. Rather, the {\it vev} $v_{\chi}$  depends on the mass of the DM, Higgs mass, and the SM-BSM Higgs mixing angle $\theta$. The neutrino mass constraint, as we will derive, would be highly dependent on additional parameters. The constraint on the Dirac Yukawa is the same as Eq.~\ref{eq:caseAN12}, 
	  The mass of $N_1$ state (i.e., $M_{N_1}$) however gets additional contribution $c^{\prime}_{11} v^2_{\Phi}/\Lambda$ due to the $N_3N_3\Phi\Phi$ term. 
	\begin{eqnarray}
	 M_{N_1}= ({c_{11} v_{\chi}^2 + c_{11}^\prime v_{\Phi}^2})/{\Lambda}
	 \label{eq:massn1}
	 \end{eqnarray}
	The cut-off scale $\Lambda$ can be written in terms of the mass of the DM, 
	\begin{eqnarray}
	\Lambda=\frac{c_{33}}{M_{N_3}} (v^2_{\chi}+\beta_3 v^2_{\Phi}),
	\label{eq:lambdab1}
	\end{eqnarray}
	where $\beta_3=c'_{33}/{c_{33}}$. We combine different constraints from Eq.~\ref{eq:caseAN12},~\ref{eq:caseBconstraint},~\ref{eq:vchicaseB1},~\ref{eq:massn1},~\ref{eq:lambdab1} in  Fig.~\ref{fig:case-B1}, where we show the variation of the Dirac coupling $y$ in $ M_{N_3}-c_{11}$ plane. For this, we choose a light neutrino mass $m_\nu = 0.05$ eV, a Higgs mixing $\sin\theta=0.1$, $M_{H_2}=250$ GeV (left panel), and $1.1$ TeV (right panel). Moreover  we assume $c_{11}^\prime = 1$ and $c_{33}^\prime=2.5 \times 10^{-6}$. 
	We have checked that for this choice of parameters the coupling $c_{33}$ which dictates $M_{N_3}$, is perturbative	in the entire  region. We also display the region where $v_\chi$ is in the range 1-10 TeV. In Fig.~\ref{fig:case-B1},  this is shown as the vertical green band as $v_{\chi}$ depends on $M_{N_3}$ but not on $c_{11}$. The cut-off scale also increases with $M_{N_3}$, we checked~(using Eq.~\ref{eq:caseBconstraint},~\ref{eq:vchicaseB1}) that at the boundary of the green band $\Lambda=(1, \ 2)\times10^6$ GeV for left panel, and $\Lambda=(7, \ 14)\times10^5$ GeV for the right panel. 
\paragraph{}It is evident from Fig.~\ref{fig:case-B1}, that the choice of a large DM mass, $M_{N_3}$, together with a larger  $c_{11}$  demands a  larger coupling $y$ after imposing the light neutrino mass and relic density constraints. In the entire region the seesaw condition $M_R > M_D$ is satisfied. The red and blue lines represent the lifetime of $N_{1,2}$ as 1 sec and $10^{-2}$ sec, respectively for the left and for the right panel. The region enclosed by the red dashed line in the right plot corresponds to $\tau(N_1)>1$ sec.

 \begin{figure}[h]
 	\centering
 \includegraphics[width=0.7\textwidth]{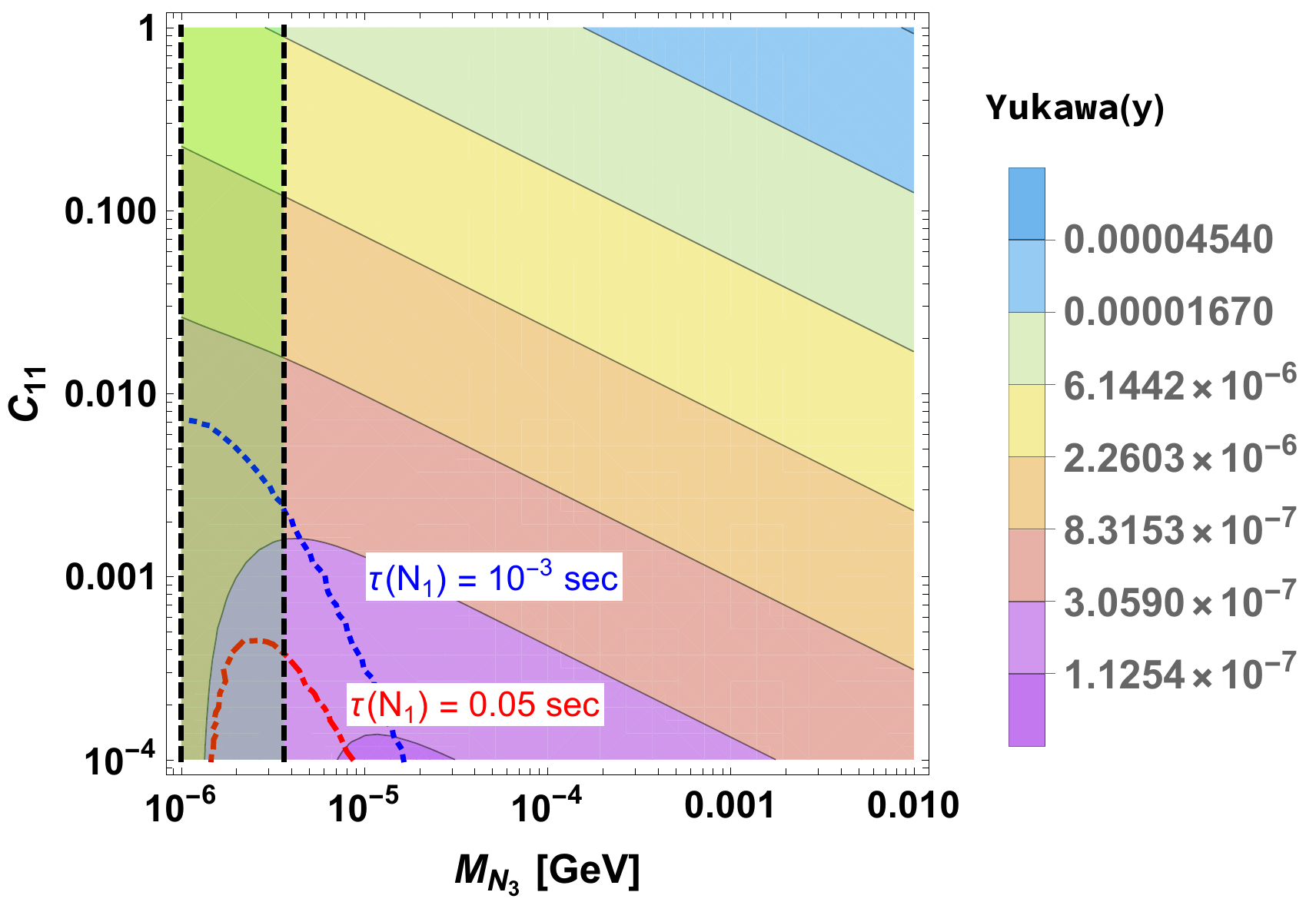}			
 	\caption{The figure corresponds to {\it Scenario-IIb}, and represents the  variation of the Yukawa coupling $y$ w.r.t the variation of the DM mass $M_{N_3}$ and the Yukawa coupling $c_{11}$.  Contours of $\tau(N_{1,2})=0.001$ (blue) and
 			0.05 (red) are displayed. The vertical black lines correspond to
 			$v_\chi=3$ TeV (left) and 10 TeV (right).}
 	\label{case-B2}
 \end{figure}

%%%%%%%%%%%%%%%%%%%%%%%%%%%%%%%%%%%%%%
 (b) The other scenario is where  DM is  produced from the SM Higgs. This can be realised for a suppressed $N_3 N_3 H_2$ coupling, we will consider the limit where this coupling is zero leading to the following constraint,
%This can be realised if the  $N_3 N_3 H_2$ coupling is zero. This scenario gives the following constraint,
\begin{eqnarray}
 \tan \theta= - \frac{c_{33} v_{\chi}}{c^{\prime}_{33} v_{\Phi}}.
\label{eq:caseB2b}
\end{eqnarray} 
	Using Eq.~\ref{eq:caseB2b},  $\tilde{\lambda}_1$ can be simplified to the form, 
	\begin{eqnarray}
	 \tilde{\lambda}_1=  \frac{- 2 M_{N_3}}{(v_{\chi} \sin \theta - v_{\Phi} \cos \theta)}
	\label{eq:NNHcaseB2}
	\end{eqnarray}
	Using  above coupling in  Eq.~\ref{eq:lambda} we get,	
	\begin{eqnarray}
	v_{\chi}={\color{blue}\pm}\frac{1.22\times 10^{12}  \ M_{N_3}^{3/2} }{ M_{H_1}^{1/2} \sin\theta} + \frac{ v_{\Phi}\cos\theta}{\sin\theta}
	\label{eq:vchicaseB2}
	\end{eqnarray}
	{In the above, the $+ \text{ and} -  \text{ sign correspond to  } (v_{\chi} \sin \theta - v_{\Phi} \cos \theta)  > 0  \text{ and} < 0$, respectively.}
 
 The constraint on the Dirac Yukawa in this case remains as in Eq.~\ref{eq:caseAN12}, where the cut-off scale $\Lambda$ and the mass of the DM are related by Eq.~\ref{eq:lambdab1}. In Fig.~\ref{case-B2} we plot the Dirac Yukawa as a function of $c_{11}$ and $M_{N_3}$. As before we consider the parameter $\sin \theta =0.1$. Additionally, we consider $c^{\prime}_{11}=1, c^{\prime}_{33}=-10^{-6}$.  The Yukawa coupling $c_{33}$ varies with $M_{N_3}$, and is perturbative in the entire range of $M_{N_3}$. The seesaw condition $M_D<M_{R}$ is satisfied in the entire parameter space. The green band bounded by black dashed lines represent the variation of $v_{\chi}$ ($\Lambda$) between 3 TeV ($2\times 10^4$ GeV) and 10 TeV ($5\times10^4$ GeV), from left to right. The lifetime of $N_{1,2}$ is less than 1 sec in the entire range. For illustration, the blue and red dashed lines indicate lifetime of $N_{1,2}$ as 0.001 sec, and $0.05$ sec, respectively.
  %\end{itemize}
  \begin{figure}[h]
  	\centering
  	\includegraphics[width=0.5\textwidth]{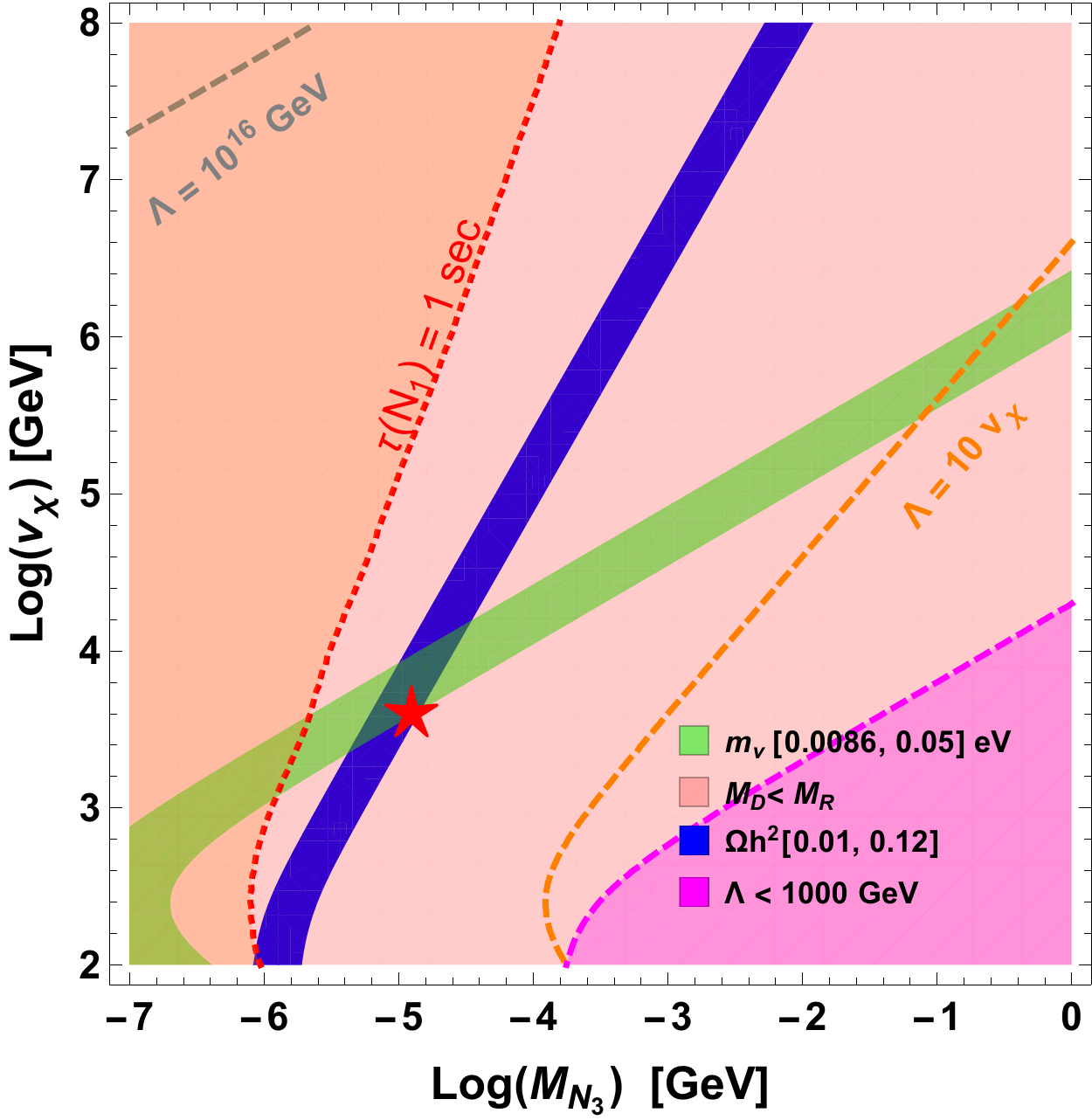}
  	\caption{ Similar as the Fig.~\ref{case-AandB}, but for {\it Scenario-II}. See text for more details.
  	}
  	\label{case-B}
  \end{figure}
  \paragraph{}
  We also consider the generic scenario where both the $H_{1}\to N_3 N_3$ and $H_2 \to N_3 N_3$ contribute to the relic density. In Fig.~\ref{case-B}, we show different constraints in the $v_{\chi} - M_{N_3}$ plane. The blue band represents the total contribution from 
 $H_1 \to N_3 N_3$ and $H_2 \to N_3 N_3$ which varies in the mentioned range. The green band represents the constraint from light neutrino mass. While the relic density constraint does not depend on the Yukawa $y$, the latter depends on few additional parameters. See Table.~\ref{tab:tab1} for the details of the input parameters. Similar to {\it Scenario-I}, we represent   $\tau(N_{1,2}) =1$ sec by red line, cut off scale $\Lambda=10 v_{\chi}$ by orange line. The point represented by a red star mark in this plot, corresponds to the star
 point shown in the left panel of Fig.~\ref{fig:case-B1}, representing the same benchmark point. Similar to the previous scenario {\it Scenario-I}, a higher {\it vev} $v_{\chi}$ is required to satisfy relic abundance for a heavier DM mass. This happens, as for both these two scenarios, the DM mass is governed by the {\it vev} $v_{\chi}$, which also governs the 
 $H_{1,2} \to N_3 N_3$ coupling, and hence the DM production. Therefore, a TeV scale $v_{\chi}$ together with  a TeV scale BSM Higgs with mass $M_{H_2}$ demand that the DM mass in this case can be at most $M_{N_3} \sim \textrm{few}$ KeV. We will see in the next section, how addition of  a bare mass term in the Lagrangian relaxes this strong correlation.
 
%%%%%%%%%%%%%%%%%%%%%%%%%%%%%%%%%%%%%%%%%%%%%%%%%%%%%%%%%%%%%%%%%%%%%%%%%%
\subsection{\it Scenario-III} 
%%%%%%%%%%%%%%%%%%%%%%%%%%%%%%%%%%%%%%%%%%%%%%%%%%%%%%%%%%%%%%%%%%%%%%%%%%%
The DM phenomenology changes if a bare mass term for the FIMP DM $N_3$ state is being added to the Lagrangian. In this case, the tight correlation between the DM mass and {\it vev} of the gauge singlet scalar relaxes.  Adding a bare mass term $M_B$ for $N_{1,2,3}$, the Lagrangian has the following form,
\begin{eqnarray}
\mathcal{L}_{eff} &= & \frac{c_{ij}}{\Lambda}  N^T_i C^{-1} N_j \chi^2+ \frac{c'_{ij}}{\Lambda}  N^T_i C^{-1} N_j \Phi^{\dagger} \Phi \nonumber \\  && +\frac{ Y_{ij}}{\Lambda} \bar{L}_i N_j H \chi+ M_{B} N^T_i C^{-1} N_i +h.c
\label{eq:caseC}
\end{eqnarray}
In the above, $M_B$ is a diagonal mass matrix $M_B=(M_{B_1}, M_{B_2}, M_{B_3})$, that represents the bare mass term for RHNs. 
The RHN mass matrix of $N_{1,2}$  can be written as follows, %  Neutrino mass constraint 
  \begin{eqnarray}
  (M_R)_{\alpha \beta} = \frac{c_{\alpha \beta} v_{\chi}^2}{\Lambda}+ \frac{c'_{\alpha \beta} v_{\phi}^2}{\Lambda}+(M_B)_{\alpha \beta}~~ (\alpha, \beta=1,2).
  \label{eq:massRHNC}
  \end{eqnarray}
  Hence, the DM mass is 
  \begin{eqnarray}
  M_{N_3} = \frac{c_{33} v_{\chi}^2}{\Lambda}+ \frac{c'_{33} v_{\phi}^2}{\Lambda}+M_{B_3}.   \label{eq:masssce3DM}
  \end{eqnarray}
 Note that, for  $c_{33}$ and $c^{\prime}_{33} $ of $\mathcal{O}(1)$, the mass of $N_3$ can primarily be governed by  $M_{B_3}$, and the other two terms can provide sub-dominant contributions.
 The Dirac mass matrix  have the same form as in previous scenarios,~Eq.~\ref{eq:MDandmRsc1}, and as before, $M_N \sim M_R$. 
  The light neutrino mass matrix has the following expression,
   \begin{eqnarray}
 m_{\nu} &\sim&  - \frac{1}{\Lambda} \frac{{Y} v^2_{\Phi} v_{\chi}^2 Y^T}{c_{11} v_{\chi}^2+ c'_{11} v_{\phi}^2+ \Lambda M_{B_1}},  
  %M_N &\sim&  \frac{c_0}{\Lambda} v^2_{\chi} (1+\frac{v^2_{\Phi}}{v^2_{\chi}})
\label{eq:massneuC}
\end{eqnarray}
where we again consider $M_{N_1}=M_{N_2}$. 
\begin{table}[b]
\centering
	\begin{tabular}{|l|l|l|l|l|l|}
		\hline
		\multicolumn{6}{|l|}{\hspace{3 cm }{\it Scenario-III}} \\ \hline
		$M_{H_2}$  &$\sin\theta$  &$y$  & $c_{33}\ (=c^\prime_{33})$ & $ M_{N_3} -M_{B_3}$ & $M_{N_1}$ \\ \hline
		$250$ GeV  & $0.1$ & $1$ & $10^{-4}$ &$10^{-8}$ GeV &$4 M_{N_3}$  \\ \hline
	\end{tabular}
	\caption{{The  parameters relevant for {\it Scenario-III}, Fig.~\ref{case-C}.}}
	\label{tab:tab2}
\end{table}
\begin{figure}[b]
	\centering
	\includegraphics[width=0.5\textwidth]{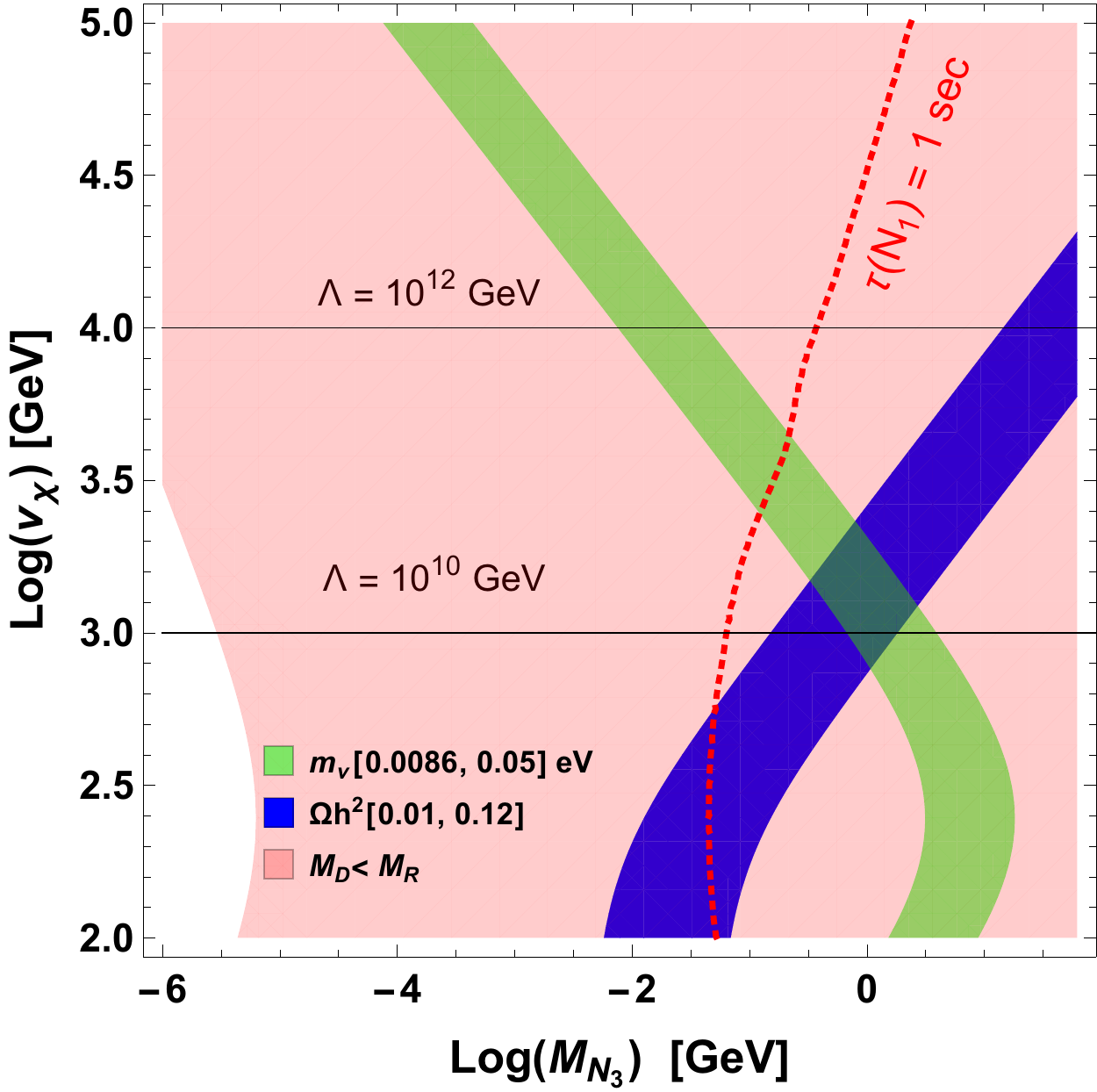}
	\caption{\small{{\it Scenario-III}: Constraints from the DM relic density
			(blue band) and eV scale light neutrino mass (green band) in the $M_{N_3}-v_{\chi}$ plane. The values of other free parameters are listed in Table~\ref{tab:tab2} . Here we include the decay of $H_{1,2}$ for DM production.	}}
	\label{case-C}
\end{figure}
%%%%%%%%%%%%%%%%%%%%%%%%%%%%%%%%%%%%%%%%%%%%%%%%%%%%%%%%%%%%%%%%
\noindent\paragraph{\bf DM production-} In this scenario, the DM can be produced from $H_{1,2} \to N_3 N_3$ decays. The interaction vertex of $N_3$ with $H_{1,2}$ states have the same form as given in Eq.~\ref{eq:DMHiggsinteractionB1}, and Eq.~\ref{eq:DMHiggsinteractionB2}. As in previous scenarios we
consider two extreme cases where   a) $\tilde{\lambda}_1$=0  and the DM is
produced from $H_2$ decay. b) $\tilde{\lambda}_2=0$ and the DM is produced
from $H_1$ decay.
\begin{enumerate}
	\item[($a$)] The condition $\tilde{\lambda}_1=0$ together with Eq.~\ref{eq:caseBconstraint} leads to  the $N_3N_3H_2$ coupling 
\begin{eqnarray} \tilde{\lambda}_2=  \frac{ 2 (M_{N_3}-M_{B_3})}{(v_{\chi} \cos \theta + v_{\Phi} \sin \theta)}.
\label{eq:NNHcasec1}
\end{eqnarray}
  and to the constraint on the {\it vev} $v_\chi$ 
  \begin{eqnarray}
v_{\chi}=\frac{1.22\times 10^{12}  \ (M_{N_3}-M_{B_3})^{3/2} }{ M_{H_2}^{1/2} \cos\theta} - \frac{ v_{\Phi}\sin\theta}{\cos\theta}.
\label{eq:vchicasec1}
\end{eqnarray}

	\item[($b$)] The condition $\tilde{\lambda}_2=0$ together with Eq.~\ref{eq:caseB2b} leads to  the $N_3N_3H_2$ coupling 
\begin{eqnarray} \tilde{\lambda}_1=  -\frac{ 2 (M_{N_3}-M_{B_3})}{(v_{\chi} \sin \theta - v_{\Phi} \cos \theta)}.
\label{eq:NNHcasec2}
\end{eqnarray}
and to the constraint on the {\it vev} $v_\chi$ 
\begin{eqnarray}
v_{\chi}=\pm\frac{1.22\times 10^{12}  \  (M_{N_3}-M_{B_3}) M_{N_3}^{1/2} }{ M_{H_1}^{1/2} \sin\theta} + \frac{ v_{\Phi}\cos\theta}{\sin\theta}.
\label{eq:vchicasec2}
\end{eqnarray}
	\end{enumerate}
For both the scenarios $a$ and $b$, it is evident  that the factor $M_{N_3}-M_{B_3}$
in the numerator of Eq.~\ref{eq:vchicasec1} and Eq.~\ref{eq:vchicasec2} will have impact on the tight correlation between $v_\chi$ and $M_{N_3}$,  found in {\it Scenario-I-II}.

\paragraph{} In general both $H_1\to N_3 N_3$ and $H_2\to N_3 N_3$ can contribute to the relic
 density. With the choice of input parameters given in Table~\ref{tab:tab2}, the constraints
 on $M_{N_3}$ and $v_{\chi}$ are shown in Fig.~\ref{case-C}.
The pink colour shaded area indicates that seesaw approximation $M_R > M_D$ is satisfied. In the blue band, DM relic density varies in the range $0.01<\Omega h^2<0.12$ from left to right. The green band represents the constraint from light neutrino mass while
the red dashed line  corresponds to the contour $\tau(N_{1,2})=1$ sec. 
The horizontal lines represent the cut-off scale $\Lambda=10^{10}$ and $10^{12}$ GeV. From Fig.~\ref{case-C} it can be seen that the region compatible with both the relic
density and the neutrino mass constraints corresponds to $v_{\chi} \sim \cal O$(TeV)  and $M_{N_3}\sim \cal O$(GeV). In this scenario it is therefore natural to have a BSM Higgs at the TeV scale for a coupling $\lambda_2 \sim{\cal O}(1)$.
In Section~\ref{collider} we will consider this scenario and explore the collider signature of a TeV scale BSM Higgs. 
%%%%%%%%%%%%%%%%%%%%%%%%%%%%%%%%%%%%%%%%%%%%%%%%%%%%%%%%%%%%%%%%%%%%%%%%%%%%%%%%%%%%%%%%
\section{DM production : decay vs annihilation \label{decvsannhi}} 
In the previous sections we considered only decay contributions of the SM and BSM Higgs  in the relic density.  This is justified for a not too large reheating temperature $T_R$~\cite{Hall:2009bx}. In this section we deviate from this assumption and allow for a high $T_R$. Thus we obtain larger contributions from annihilation channels as well. To determine the co-moving number density of $N_3$, we need to solve the following Boltzmann equation,
\begin{eqnarray}
&&\dfrac{dY_{N_3}}{dz} = \dfrac{2 M_{pl}}{1.66 M_{H_2}^{2}}
\dfrac{z \sqrt{g_{\star}(z)}}{g_{s}(z)}\,\,\Bigg[\sum_{i = 1,\,2}
\langle\Gamma_{H_{i} \rightarrow  N_{3} N_{3}} \rangle
(Y_{H_i}^{eq} - Y_{N_3})\Bigg]\nn \\
&&~~~~~~~~~
+\dfrac{4 \pi^{2}}{45} \dfrac{M_{pl} M_{H_2}}{1.66}
\dfrac{\sqrt{g_{\star}(z)}}
{z^{2}}\,\,\Bigg[\sum_{x = W,Z,h_{1},h_{2},f}
\langle\sigma {\rm v}_{x\bar{x}\rightarrow N_{3} N_{3}}\rangle (Y_{x}^{eq\,\,2} - Y_{N_{3}} ^{2})\Bigg]
\label{BE}
\end{eqnarray}

%where $g_{s}(z)$, $g_{\rho}(z)$ are the entropy and matter d.o.f of the Universe and 
where $z=M_{H_2}/T$ and $g_{*}(z)$ depends on $g_{s}(z)$, $g_{\rho}(z)$ in the following way \cite{Gondolo:1990dk}, $\sqrt{g_{*}(z)} = \frac{g_{s}(z)}{\sqrt{g_{\rho}(z)}} \left( 1 - \frac{1}{3} \frac{d\, {\rm ln}\, g_{s}(z)}{d\, {\rm ln}\, z} \right)\,.$
 $Y_{A}$ is the co-moving number density of $A$, $M_{pl}$ is the Planck mass and
the quantity inside $<...>$ corresponds to the thermal average of decay rate and annihilation cross-sections. From the co-moving number density obtained by solving the Eq.\,\ref{BE}, one  can determine the relic density of DM by  
following the expression,
\begin{eqnarray}
\Omega_{N_3} h^{2} = 2.755 \times 10^{8}\, \left(
\dfrac{M_{N_3}}{\rm GeV}\right)\,
Y_{N_3} (T_{0})\,,
\label{rel-den}
\end{eqnarray}
We perform our numerical simulation using micrOMEGAs5.0  \cite{Belanger:2018ccd} after implementing the model file in Feynrules \cite{Alloul:2013bka}. 

\begin{figure*}[b]
	\centering
	\includegraphics[width=0.45\textwidth]{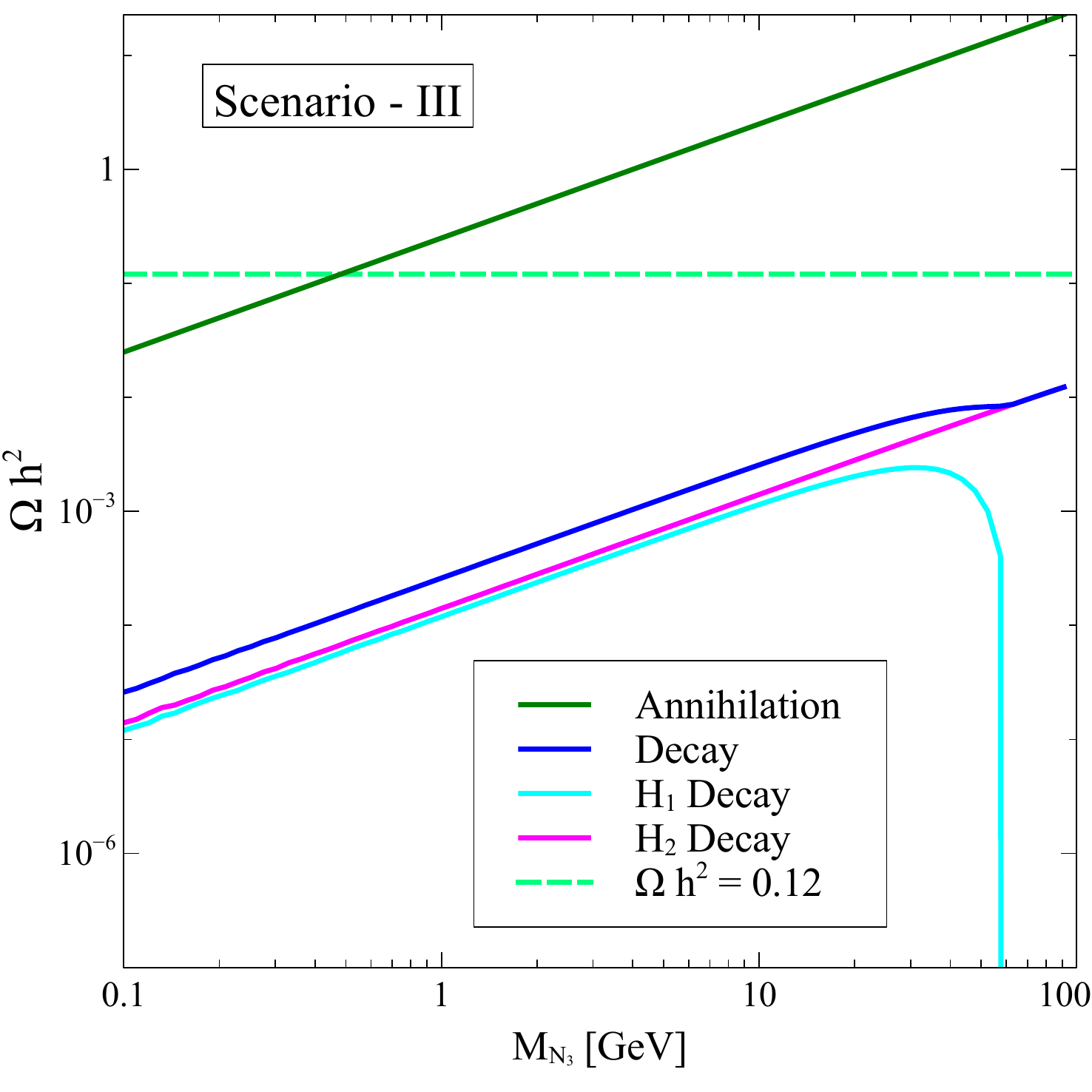}
	\includegraphics[width=0.45\textwidth]{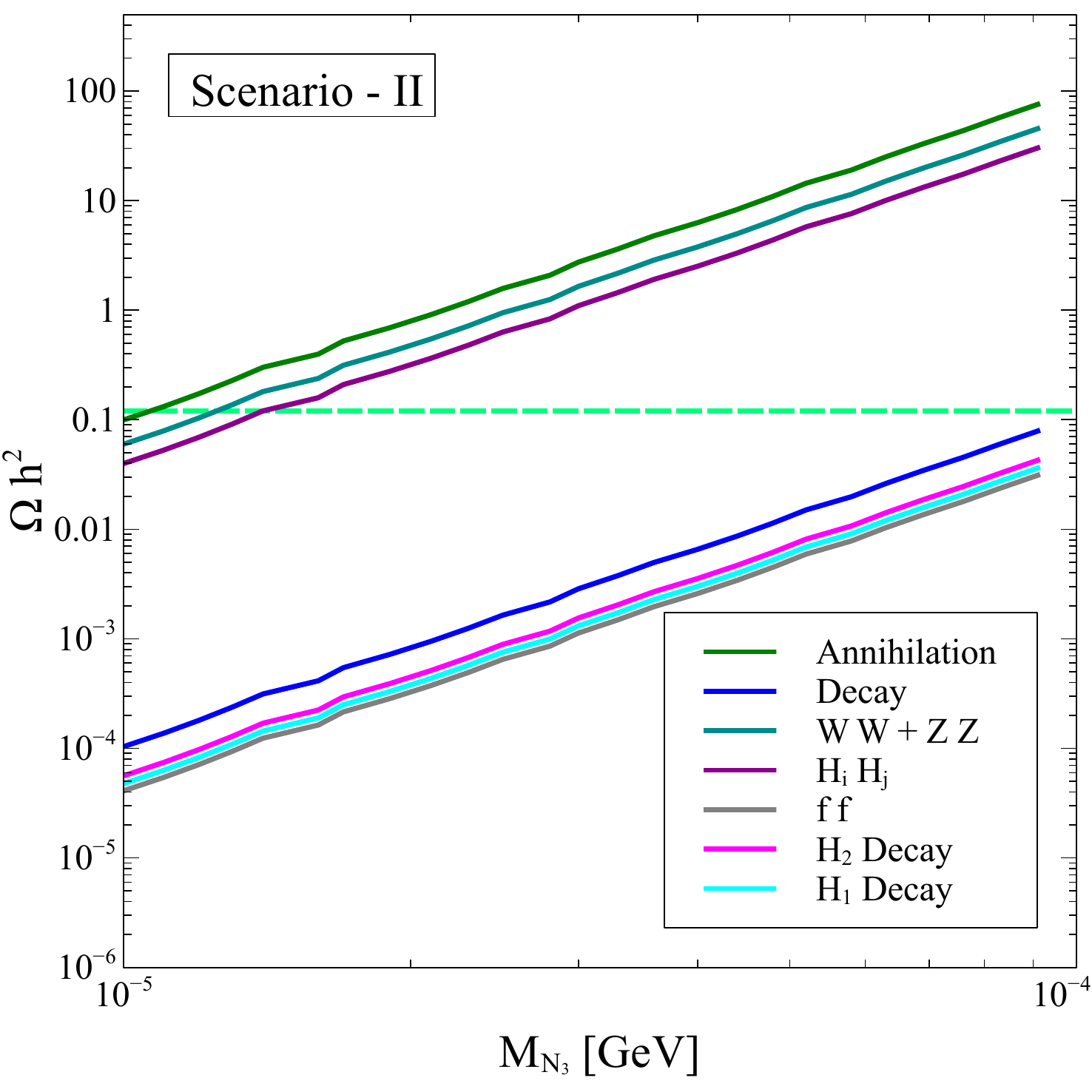}
	\caption{\small{Left~(right) panel : $\Omega h^2$ vs the mass of DM for {\it Scenario-III} ({\it Scenario-II}) with $\sin \theta = 0.1$, $\Lambda = 10^{11}$ GeV, $M_{H_2} = 1100$ GeV, 
	$M_{N_{1,2}} = 5 M_{N_3}$, $M_{N_3} - M_{B_3} = 10^{-8}$ GeV, 
	$v_{\chi} = 3$ TeV (left panel), $M_{B_3} = 0$, $v_{\chi} =15$ TeV (right panel). The green-dashed horizontal line represents $\Omega h^2=0.12$ \cite{Aghanim:2018eyx}. }}
	\label{line-plot-1}
\end{figure*}

In the left  and right panel  of Fig.~{\ref{line-plot-1}}, we show the variation of the DM relic density with its mass for {\it Scenario-III } and {\it Scenario-II}, respectively. We keep the other parameters fixed as shown in the caption. For these figures, we choose a high value of the reheating temperature, $T_{R} = 10^{9}$ GeV. From both the figures, one  can see that the DM relic density
increases with its mass $M_{N_3}$ as also evident from Eq.~\ref{rel-den}. In the left panel, there is  a sharp fall in  the 
$H_1$ decay contribution at $M_{N_3} = \frac{M_{H_1}}{2}$ when the $H_1$ decay
to DM is not kinematically allowed. Note that for large $T_R$ the
annihilation contribution is larger than the decay contribution by
several orders of magnitude. This occurs as   relic density for annihilation 
(from $WW/ZZ\to N_3N_3,\ H_iH_j\to N_3N_3$) is proportional to the value of the reheating
temperature as shown in the Appendix, see Eq.~\ref{dependence-yuv-TR}.   
\begin{figure*}[t]
	\centering
	\includegraphics[width=0.45\textwidth]{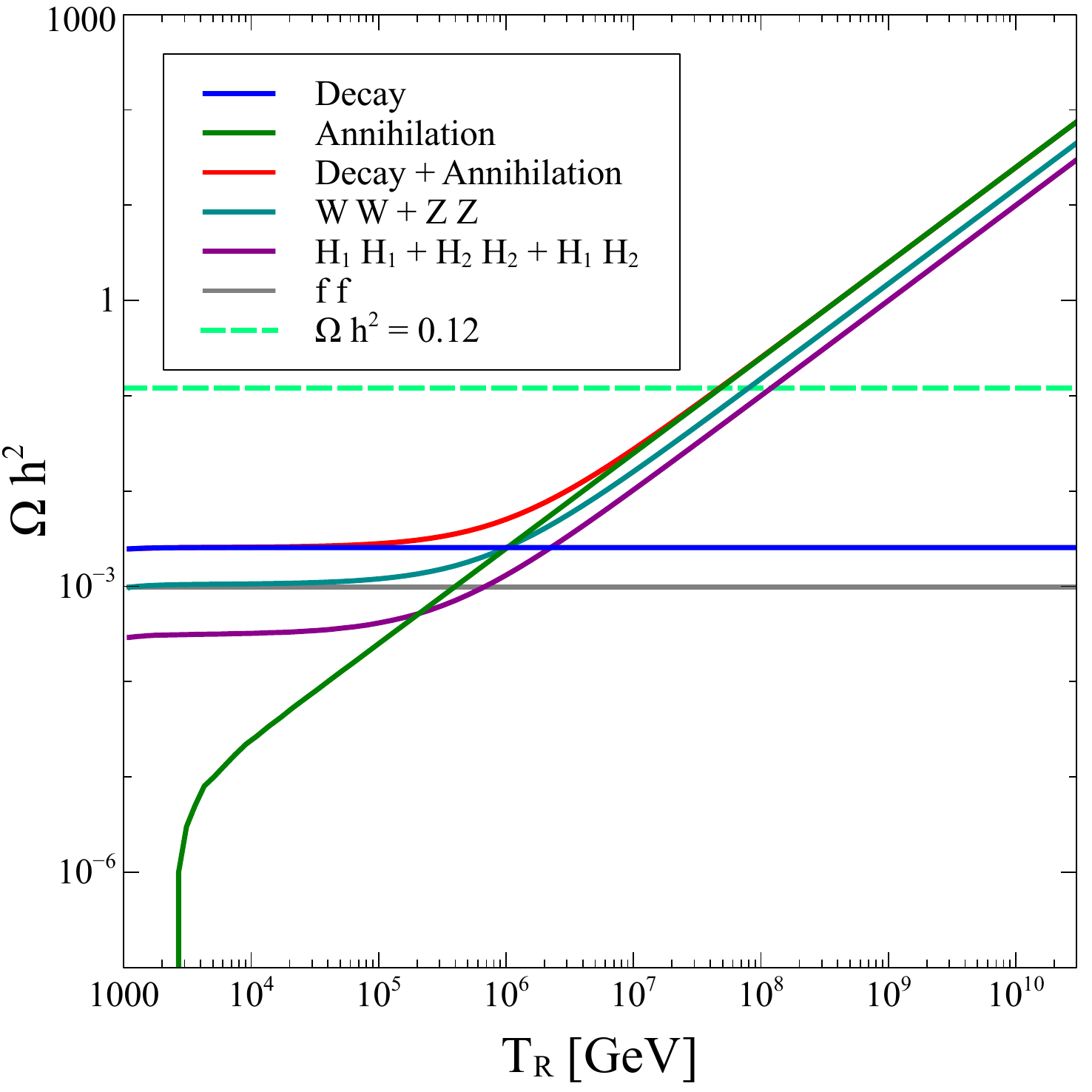}
	\includegraphics[width=0.45\textwidth]{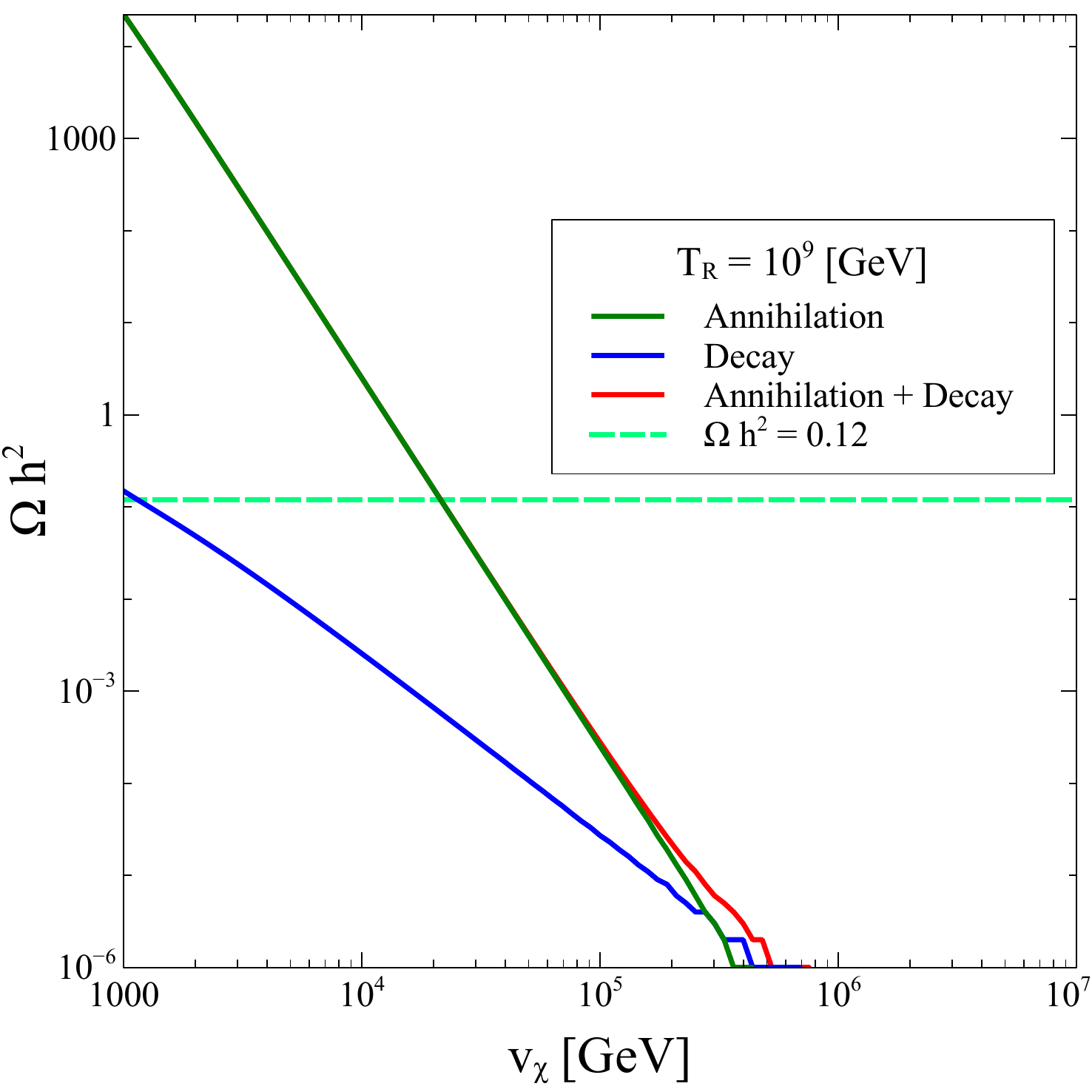}
	\caption{\small{Left panel: Relic density vs reheating temperature for {\it Scenario-III}. Right panel:    
	variation of relic density vs {\it vev} of the singlet scalar for the same scenario. 
	Other parameters kept fixed at $\sin \theta = 0.1$,
	$\Lambda = 10^{11}$ GeV, $M_{H_2} = 1100$ GeV, $M_{N_{1,2}} = 5 M_{N_3}$, $M_{N_3} - M_{B_3} = 10^{-8}$ GeV and 
	$v_{\chi} = 3000$ GeV. The green dashed horizontal line represents the experimental  constraint on DM relic density \cite{Aghanim:2018eyx}. }}
	\label{line-plot-2}
\end{figure*}

We also show the variation of relic density with the reheating temperature $T_R$ (shown in the left panel), and the {\it vev} $v_{\chi}$ (shown in the right panel)  in Fig.~\ref{line-plot-2}. Moreover, the individual contributions from decay and annihilation processes
to the total relic density have also been shown in both the panels. In the left panel, we can see that the decay contribution does not depend on the reheating temperature, $T_{R}$, but annihilation contribution strongly depends on the reheating temperature when $T_{R} \gsim 10^{6}$ GeV. As we have shown  in Eq.\,\ref{dependence-yuv-TR} of the Appendix, and also have been discussed in  \cite{Hall:2009bx},   the annihilation contribution depends on the reheating temperature and linearly grows with it. This is visible in Fig.~\ref{line-plot-2}, where for large $T_R$ annihilation contributions increase. This 
occurs because of the presence of $d=5$  $H_{i} H_{j} N_{3} N_{3}$ ($i,j = 1,2$) operators. We also show in the Appendix that $WW/ZZ\to N_3N_3$ exhibits similar  feature. 
 Furthermore, as evident from Eq.~\ref{dependence-ff-yuv-TR}, the contribution from the annihilation process $f \bar{f} \rightarrow N_{3} N_{3}$ is dominated by the low scale physics. For a very high $T_R$ this contribution becomes almost independent of the  variation of $T_R$.
 In the right panel of  Fig.~\ref{line-plot-2}, the variation of relic density with the \textit{vev} $v_\chi$ has been shown. 
One can see that both decay and annihilation contributions fall linearly with the increase of the $v_{\chi}$. This  can be explained easily, as  the   coupling of DM with the Higgs  is inversely proportional to the square of the {\it vev} {\it i.e.} 
$\tilde{\lambda}_{1,2} \propto \frac{1}{v^2_{\chi}}$. In addition, the $WW \to N_3N_3, ZZ \to N_3N_3$ contributions also depend on the coupling $\tilde{\lambda}_{1,2}$. 
Moreover the contact interaction $H_i H_i \to N_3 N_3$ depends on $c_{33}/\Lambda$, which also varies as $1/v_{\chi}$. Therefore, as  the {\it vev} increases,  the relevant  coupling becomes smaller, resulting in a reduced production of DM.

%%%%%%%%%%%%%%%%%%%%%%%%%%%%%%%%%%%%%%%%%%%%%%%%%%%%%%%%%%%%%%%%%%%%%%%%%%%%%%%%%%%%%%%%
\subsection{The parameter space of {\it Scenario-III } \label{decayvsannihilation}}
In the previous section, we have evaluated  the decay and annihilation contributions to the relic density for specific benchmark points. 
In this section we vary all the free
parameters of {\it Scenario-III}  in a wide range and present the results in
the form of scatter plots. The model parameters are varied in the
following range,
\begin{eqnarray}
\label{eq:scatpoints}
200\,\,{\rm GeV}\,\, < &M_{H_2}& < 3000\,\,{\rm GeV} \\ \nn
10\,\,{\rm GeV}\,\, < &M_{N_3}& < 100\,\,{\rm GeV} \\ \nn
10^{-3} < &\theta& < 10^{-1} \\ \nn
1000\,\,{\rm GeV}\,\, < &v_{\chi}& < 10000\,\,{\rm GeV} \\ \nn
200\,\,{\rm GeV}\,\, < &T_{R}& < 10^{9}\,\,{\rm GeV} \\ \nn
10^{9}\,\,{\rm GeV}\,\, < &\Lambda& < 10^{14}\,\,{\rm GeV}\,. 
\end{eqnarray}
To accommodate  $H_2$ at the TeV scale together with $v_\chi\sim$TeV, the
	bare mass term of $N_3$ has to dominate its physical mass $M_{N_3}$, thus we impose
	$M_{N_3}-M_{B_3} \sim \mathcal{O}(10^{-8})$ GeV.
	In our scan we require that $N_3$ contribute to at least $10\%$ of the total
	DM, thus we impose that its relic density falls within the range 
\begin{eqnarray}
0.01 < \Omega h^2 < 0.1211.
\label{relic-density-bound}
\end{eqnarray}  

\begin{figure*}[t]
	\centering
		\includegraphics[width=0.45\textwidth]{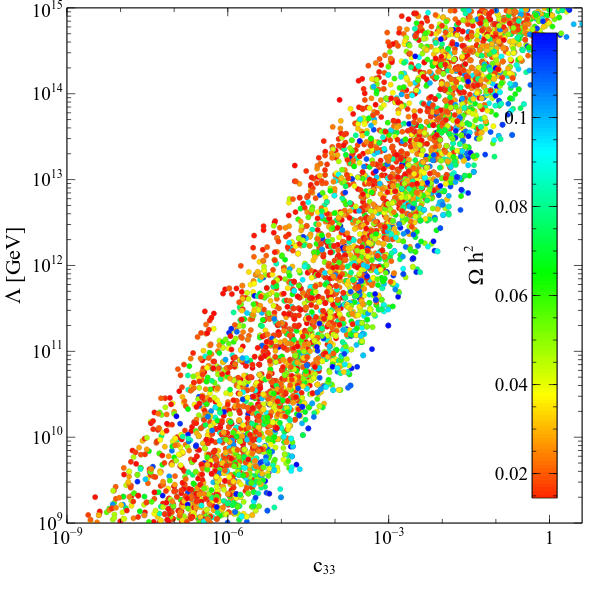}
		\includegraphics[width=0.45\textwidth]{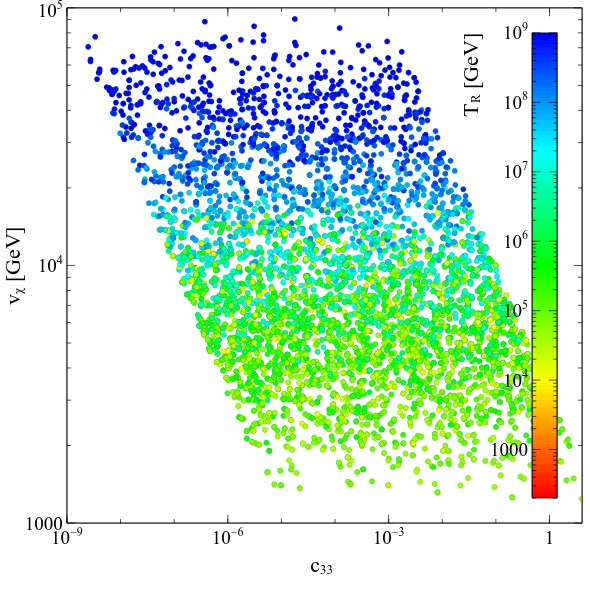}
		\caption{\small{Left panel: variation of relic density of $N_3$  in the $c_{33} - \Lambda$ plane.  Right panel:  variation of $T_R$  in the $c_{33}-v_{\chi}$ plane. We consider the same constraint as in the left panel.}}
	\label{scatter-plot-1}
\end{figure*}

\begin{figure*}[b]
	\centering
	\includegraphics[width=0.45\textwidth]{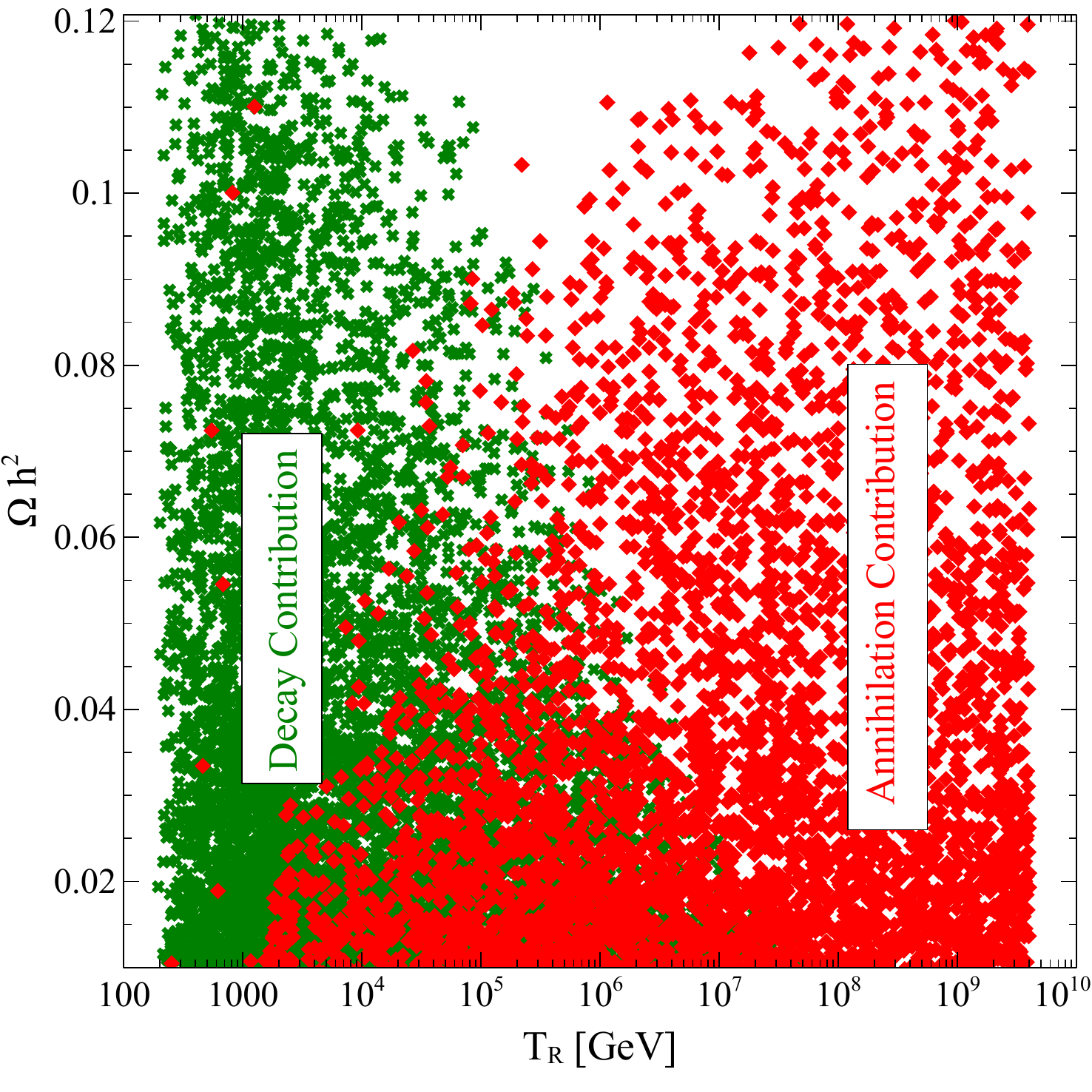}
	\includegraphics[width=0.45\textwidth]{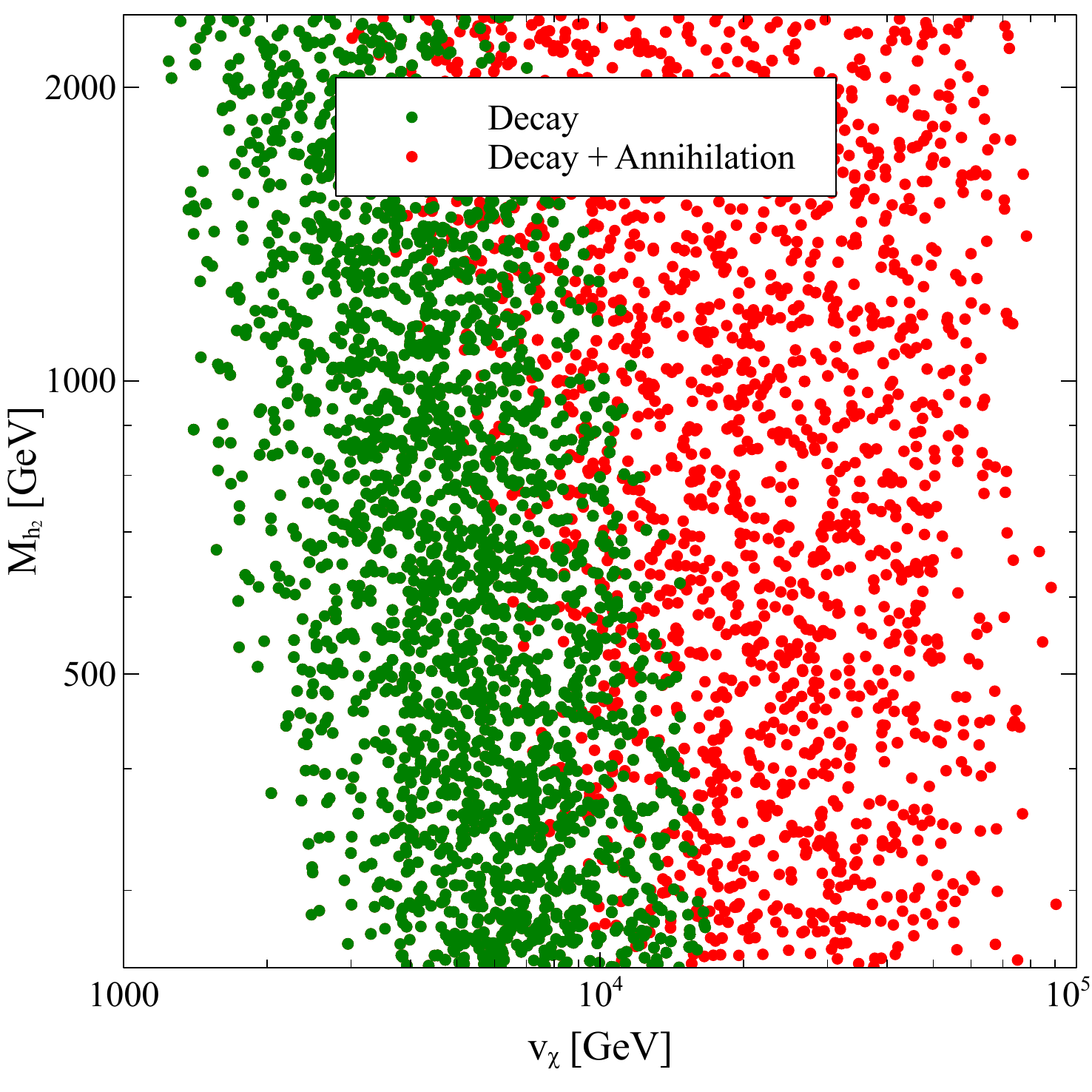}
	\caption{\small{ Left panel:  Variation of relic density of $N_3$ vs $T_R$.  Right panel:  scatter plot in the $v_{\chi} - M_{H_2}$ plane.}}
	\label{scatter-plot-2}
\end{figure*}

  In Fig.~\ref{scatter-plot-1}, \ref{scatter-plot-2} we display the allowed parameter space after taking into account the constraint from Eq.~\ref{relic-density-bound}. The entire range of $M_{N_3}$ mentioned above can satisfy Eq.~\ref{relic-density-bound} with the variation of other model parameters. We did not find any strong correlation between $M_{N_3}$ and other model parameters, and hence we do not present any scatter plot for $M_{N_3}$. 

\paragraph{}In the left panel of Fig.~\ref{scatter-plot-1}, we  show the variation of relic density (in color bar) in the $c_{33} - \Lambda$ plane after satisfying Eq.~\ref{relic-density-bound}. For {\it Scenario-III},
we can express the coupling $c_{33}$ in terms of the cut off scale $\Lambda$ and the {\it vevs} $v_{\phi}, v_{\chi}$ in the 
following way,
\begin{eqnarray}
c_{33} = \frac{\Lambda (M_{N_3} - M_{B_3})}{v^2_{\chi} + \beta_3 v^2_{\phi}}\,.
\label{c33-expression}
\end{eqnarray} 
As evident from the above expression, there is a linear relation
between $c_{33}$ and $\Lambda$. Therefore, as  we increase $\Lambda$, $c_{33}$
also increases. This is clearly visible from the figure shown in the left panel. 
The blue scattered points satisfy the experimentally measured DM relic density constraint \cite{Aghanim:2018eyx}. In the right panel of Fig.~\ref{scatter-plot-1}, we show the points that satisfy  Eq.~\ref{relic-density-bound} in the $v_{\chi} - c_{33}$ plane. As expected from Eq.~\ref{c33-expression}, since $v_\chi\gg\beta_3 v_\phi$, $c_{33}$ is inversely proportional to $v_\chi^2$. Furthermore, right panel of Fig.~\ref{scatter-plot-1} shows that $T_R$ increases linearly with $v_\chi$. This can be understood as follows. We have seen in  Fig.~\ref{line-plot-2} that the relic density (when dominated by the annihilation contribution) increases with the reheating temperature at large $T_R$, and decreases  as $1/v_\chi$. Therefore, for a given value of the DM relic density,  higher values of $T_R$ will be associated with larger values of $v_\chi$. The yellow points are not clearly visible, as they have been covered by the green points.

\begin{figure*}[b]
 	\centering
 	\includegraphics[width=0.45\textwidth]{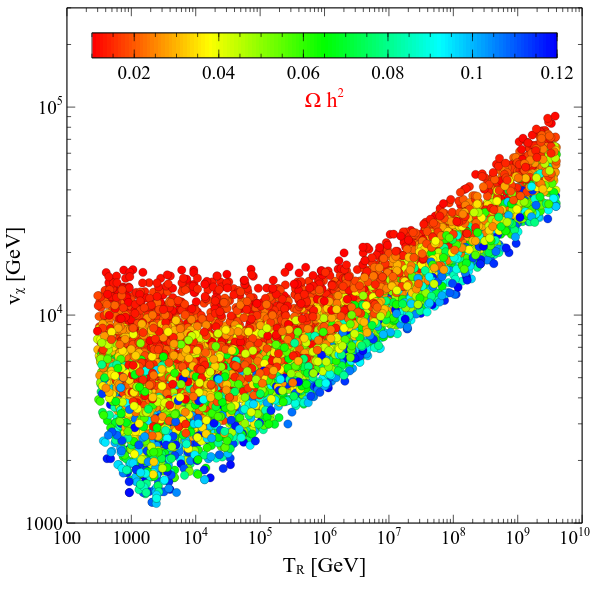}
 	\caption{Variation of relic density w.r.t variation of $v_{\chi}$ and $T_R$. The points satisfy  the mentioned DM relic density range.}
 	\label{scatter-plot-3}
 \end{figure*}

\paragraph{}
In the left panel of Fig.\,\ref{scatter-plot-2}, we  show the variation of the  decay and annihilation contributions to the relic density  with $T_R$.  For the discussion on the dependency of the relic density on $T_R$, see Section.~\ref{App:AppendixA}. In the right panel of the same plot, we show the relation between $v_\chi$ and $M_{H_2}$.  As evident from the left plot, for  reheating temperature $T_R \sim 10^5$ GeV,  the decay and annihilation contributions
are equal, while for $T_R > 10^6-10^7$ GeV,  the annihilation contributions dominate. Lower than $T_R \sim 10^5$ GeV,  decay contribution to the relic density dominates. This occurs as the 
annihilation contribution is directly proportional to the reheating temperature as has been explained before. In generating the scatter plots both for Fig.~\ref{scatter-plot-1} and Fig.\,\ref{scatter-plot-2}, we assume that the reheating temperature is greater than the masses of all the particles.
\paragraph{}In the right panel  of  Fig.~\ref{scatter-plot-2}, we  show the points in the $v_{\chi} - M_{H_2}$ plane which satisfy Eq.~\ref{relic-density-bound}. We represent the decay contribution by green points and the total 
contribution by red points. For the decay contribution, there exists an inverse correlation between $v_{\chi}$ and $M_{H_2}$. The $H_{2} N_{3} N_{3}$ coupling for this 
case takes the following form,
\begin{eqnarray}
\tilde{\lambda}_2  = \frac{2 (M_{N_3} -M_{B_3})\times 
(v_{\chi} \cos \theta + \beta_3 v_{\phi} \sin \theta)}{v^2_{\chi} + \beta_3 v^2_{\phi}}\,.
\end{eqnarray} 
The above equation together with  Eq.~\ref{eq:indomega} imply that the DM relic density decreases with the increase of both $v_{\chi}$ and $M_{H_2}$. Therefore, for a given value of the DM relic density, higher values of $v_{\chi}$ will be associated with smaller values of $M_{H_2}$. For annihilation processes the correlation between $v_{\chi}$ and $M_{H_2}$ is somewhat mild,  as the additional parameter $T_R$ plays a significant role in annihilation processes,  and we have varied $T_R$ in a wide range mentioned in Eq.~\ref{eq:scatpoints}.
 \begin{table}[b]
\centering
		\begin{tabular}{|c|c|c|c|}
			\hline
			$\sigma$ (pb) &$M_{H_2}=250$ GeV  &$M_{H_2}=500$ GeV  &	$M_{H_2}=1100$ GeV   \\ \hline
			$\sigma(p p \to H_2 \to ZZ)$ &$ <0.12$ &$<0.02 $ & $< 0.005$  \\ \hline
			$\sigma(p p \to H_2 \to VV)$  & $<1.6$ \tiny{($M_{H_2}=300$ GeV)} &$<0.2$ &$<0.008$\\ \hline
			$\sigma(p p \to H_2 \to H_1 H_1)$ &  $<0.8 \ (3)$ &$<0.15 \ (0.25)$ & $< 0.03 \ (0.04)$ \\ \hline
			$\sigma(p p \to H_2 j j  \to H_1 H_1 j j)$ & $<1$ &$<0.05$  & $<0.004$ \tiny{($M_{H_2}=1$ TeV)} \\ \hline
		\end{tabular}
	\caption{{LHC constraints on BSM Higgs production in GF and VBF channel. We follow  \cite{Aad:2020fpj} for $ZZ$ channel (ATLAS search-139 $\rm{fb}^{-1}$),  \cite{Aad:2020ddw} for $VV \ (WW+ZZ)$ channel (ATLAS search-139 $\rm{fb}^{-1}$), 
	\cite{Aad:2019uzh} for $H_1 H_1$ (ATLAS search-36.1 $\rm{fb}^{-1}$), and \cite{Aad:2020kub} for $ H_1 H_1+ j j$ channel (ATLAS search-126 $\rm{fb}^{-1}$). For $H_1H_1$ channel, the limits from CMS search-35.9 $\rm{fb}^{-1}$~	\cite{Sirunyan:2018zkk} are mentioned in bracket, that we obtained after deviding  $\sigma(p p \to H_2 \to H_1 H_1\to 4b)$ by $Br^2(H_1\to b \bar{b})$.  }}
	\label{tab:tablhc}
\end{table}

\begin{table}[b]
	\centering

\end{table}

Finally in Fig.~\ref{scatter-plot-3}, we  show the variation of the relic density w.r.t to the variation of {\it vev} $v_{\chi}$ and  the reheating temperature $T_{R}$.  As we have discussed, the coupling strength to produce DM
varies inversely with the {\it vev}, therefore for a smaller {\it vev}, the coupling $c_{33}$ increases. This results in a higher value of the $\Omega h^2$ which is represented by the blue points. We can also see for $T_{R} >10^{6}$  GeV there exist a sharp correlation between $T_{R}$ and $v_{\chi}$, which is consistent with  the right panel of Fig.~\ref{scatter-plot-1}.
%%%%%%%%%%%%%%%%%%%%%%%%%%%%%%%%%%%%%%%%%%%%%%%%%%%%%%%%%%%%%%%%%5
\section{ Collider Signature of $H_2$ \label{collider}}
Other than the SM Higgs, the model also contains a  neutral BSM Higgs, which can be probed at collider experiments. A number of LHC measurements  constrain the presence of such a heavy Higgs, and its mixing with the SM 125 GeV Higgs~\cite{Sirunyan:2018koj}.  For the collider analysis we consider BSM Higgs in the TeV mass range. We pursue the study for {\it Scenario-III}. The model signature we study  will remain the same  for {\it Scenario-I, and II} as well,  as the signature does not depend on DM mass. The LHC searches that constrain the BSM Higgs and its mixing  are - a) The SM Higgs signal strength measurement, and b) heavy Higgs searches.  

a) Higgs signal strength measurement constrains the mixing of the SM and BSM Higgs.  To evaluate this, we adopt~\cite{Sirunyan:2018koj}.  The signal strength of SM Higgs is given as, 
\begin{equation}
\mu_{h\rightarrow x x}=\frac{\sigma_{H_{1}}}{\sigma_{H_{1}}^{SM}}\frac{Br(H_{1}\rightarrow x x)}{Br^{SM}(H_{1}\rightarrow x x)}
\end{equation}
In the above,  $H_{1}\rightarrow x x$ represents   any channel where SM Higgs can decay. In our model one of the additional final states  in which SM Higgs can decay is to DM pair ($H_{1}\to N_{3}N_{3}$).  However,  the branching ratio of this channel is very small $Br(H_{1}\rightarrow N_{3}N_{3}) < 10^{-8}$. The other modes such as $H_{1}\rightarrow \nu N_{1,2}/ N_{1}N_{1}/N_2N_2$ also have small branching ratios. Therefore, for all practical purposes, these channels can be neglected. % we can neglect. 
 The branching ratio of $H_1$ decaying to  any SM final state is hence almost identical to the branching ratio in the SM, {\it i.e.,}  $Br(H_{1}\rightarrow x x)\sim Br^{SM}(H_{1}\rightarrow x x)$. The production cross-section  of $p p \to H_1$ becomes $\sigma_{H_{1}} = \cos^{2}\theta \sigma_{H_{1}}^{SM}$. Therefore, we find that  %
 the above Higgs signal strength expression takes a very simplified form,
\begin{eqnarray}
\mu_{H_{1}\rightarrow x x}\sim \cos^{2}\theta.
\label{eq:higgssignalsimple}
\end{eqnarray} 
The $\sqrt{s}=13$ TeV LHC  measurements  of the Higgs signal strength in combined channel dictates  $\mu = 1.17 \pm 0.1$~\cite{Sirunyan:2018koj}. We find, allowing a  $3\sigma$ deviation around the best fit value of $\mu = 1.17$, that the Higgs mixing angle  $\sin\theta < 0.36$. In our subsequent analysis, we consider both  a larger value of $\sin \theta=0.34$ and a small value, $\sin \theta=0.1$.  

b) A number of LHC searches constrain the production of heavy scalar resonance and its decay into various  SM final states. As discussed in the previous sections, one of the main production channels of the FIMP DM is the decay of $H_2$ to two $N_3$ states, which is dominant for a low reheating temperature.  However, due to the negligible branching ratio $Br (H_2/H_1 \to N_3 N_3 ) < 10^{-8}$, this channel is not constrained by the LHC searches for the invisible decays of Higgs. The  main decay channels of the BSM Higgs include $H_2 \to WW, ZZ, b \bar{b}, \tau^+ \tau^-,  H_1 H_1$ channels. 
	%	 In addition to the  decay $H_2 \to N_3 N_3$,  the BSM Higgs $H_2$ also decays to a number of SM states. These include $H_2 \to WW, ZZ, b \bar{b}, \tau^+ \tau^-$, and $H_2 \to H_1 H_1$ channels.
	  A number of CMS and ATLAS searches constrain the production cross-section of the BSM Higgs in gluon fusion (GF), or vector boson fusion (VBF) channels folded with the branching ratio of the $H_2$ in the above mentioned modes. In Table.~\ref {tab:tablhc}, we outline the most sensitive searches for a neutral Higgs at the LHC, where we quote the limits on $\sigma \times Br$ for  few illustrative mass points. We consider  the  searches  $p p \to H_2 \to Z Z$ \cite{Aad:2020fpj}, $p p \to H_2 \to W^+ W^-+ZZ$ \cite{Aad:2020ddw},  $p p \to H_2 \to H_1 H_1  \to 4b$ \cite{Aad:2019uzh,Sirunyan:2018zkk}, $p p \to H_2+ j j \to H_1 H_1+ j j$ \cite{Aad:2020kub}. We find that among them $p p \to H_2 \to Z Z$ \cite{Aad:2020fpj} is most constraining, in particular this channel does not
	  allow  lighter masses, $M_{H_2 }  \sim  200 $  GeV,  for larger value of $\sin \theta = 0.34 $.  Note that  this mixing angle is  allowed by Higgs signal strength
	  measurements. Such values are marginally allowed by the search  $p p \to H_2 \to VV$. On the other hand, the large mixing angle, $\sin \theta \sim 0.34$ is allowed by all the above mentioned searches when $M_{H_2} \sim 1 $ TeV. The mixing angle $\sin \theta =0.1$ which we consider for the DM analysis in the previous section is allowed for the entire range $200\,  \textrm{GeV} < M_{H_2} < \textrm{few}$ TeV. We have cross-checked our results with the results obtained from  HiggsBound \cite{Bechtle:2020pkv}.
\paragraph{} Other than the $WW,ZZ$ channels, one of the spectacular signature of a heavy BSM scalar  is the di-Higgs signal. The di-Higgs channel is in-particular  important to probe Higgs tri-linear coupling. Any deviation from the SM prediction will indicate new physics. Di-Higgs production in SM, which is non-resonant, has  extensively been studied  for LHC. There are  different  studies that analysed   $b\bar{b}\gamma\gamma$\cite{Baur:2002qd,Baur:2003gp,Azatov:2015oxa,Kling:2016lay}, $b\bar{b}W^{+}W^{-}$\cite{Dolan:2012rv, Papaefstathiou:2012qe}, $b\bar{b}\tau^{+}\tau^{-}$\cite{Dolan:2012rv,Barr:2013tda,Banerjee:2018yxy}, $b\bar{b}b\bar{b}$\cite{deLima:2014dta,Wardrope:2014kya,Behr:2015oqq,Banerjee:2018yxy},  and $b\bar{b}+E_{T}$\cite{Banerjee:2016nzb} final states. With a heavy BSM Higgs which couples to two SM Higgs, the di-Higgs production cross-section becomes large. The studies \cite{Dolan:2012ac,No:2013wsa,Martin-Lozano:2015dja,Huang:2017jws,Ren:2017jbg, Basler:2019nas, Li:2019tfd,Adhikary:2018ise,Lu:2015qqa,Alves:2019igs} focus on resonant production of a BSM Higgs and its decay to di-Higgs. Both resonant and non-resonant di-Higgs production processes have been extensively explored by CMS and ATLAS~\cite{Aaboud:2016xco, Khachatryan:2016cfa, Khachatryan:2016sey, Aad:2015xja, Aaboud:2018knk,Aaboud:2018zhh,Aaboud:2018sfw,Aaboud:2018ksn,Aaboud:2018ftw,Aaboud:2018ewm,Aad:2019uzh}. Most of the above mentioned studies focused on the resolved final states with isolated final state leptons, jets, and photons.  However, for a very heavy mass of the BSM Higgs, the produced SM Higgs will be  boosted, leading to collimated decay products. Note that,  the di-Higgs production from a  heavy Higgs with few TeV mass  is less favourable at $\sqrt{s}=13$ TeV LHC due to smaller production cross section. For non-resonant di-Higgs production at the proposed  $\sqrt{s}=100$ TeV LHC,  see~\cite{Cao:2016zob, Chang:2018uwu, Mangano:2020sao, Park:2020yps, Borowka:2018pxx}.

\paragraph{}We instead focus on resonant di-Higgs production from the decay of a heavy  BSM Higgs at  $100$ TeV collider, decaying into two SM Higgs. The branching ratio of this channel $Br(H_{2}\rightarrow H_{1}H_{1})\sim 25\%$ for $H_{2}$ mass around $1\, \rm TeV$~\cite{Banerjee:2015hoa}. We analyse the di-Higgs channel with subsequent decay of $H_1$ to $b\bar{b}$. We assume $M_{H_2} > 1$ TeV,  for which the two Higgs bosons produced from $H_2$ are moderately boosted, leading to a peak in $\Delta R$ separation between the two $b$ quarks as $\Delta R(b, \bar{b} \lesssim 0.4)$.  Instead of a resolved analysis with  four or more number of  isolated $b$ jets in the final state, we perform an analysis where we adopt a  large radius jet as the jet description, which is effective in suppressing a number of SM backgrounds. Therefore, our model signature is 
\begin{eqnarray}
%\textrm{Signal:}\, \, ~~ 
  pp \to H_2 \to H_1 H_1 \to 2 j_{\textrm{fat}}. 
\label{signalcol}
\end{eqnarray}
%\end{document}
where, each of the fatjet $j_{\textrm{fat}}$ contains two $b$ quarks appearing from Higgs decay.  
\paragraph{}To evaluate the signature, we implement the Lagrangian of this model  in  FeynRules(v2.3)~\, \cite{Alloul:2013bka} to create the UFO~\cite{Degrande:2011ua}  model files. The Event generator \texttt{MadGraph5\_aMC@NLO}(v2.6)~\cite{Alwall:2014hca} is used to generate both the signal and the background events at leading order. Generated events
 are passed through Pythia8~\cite{Sjostrand:2014zea} to perform showering and
 hadronization. Detector effects are simulated
 using Delphes (v3.4.1)~\cite{deFavereau:2013fsa}. We use FastJet~\cite{Cacciari:2011ma} for the clustering of fatjets and consider Cambridge-Achen~\cite{Dokshitzer:1997in, Wobisch:1998wt}
 algorithm,  with radius parameter $R = 1.0$.
  \begin{figure}[t]
  	\centering
  	\includegraphics[width=0.7\textwidth]{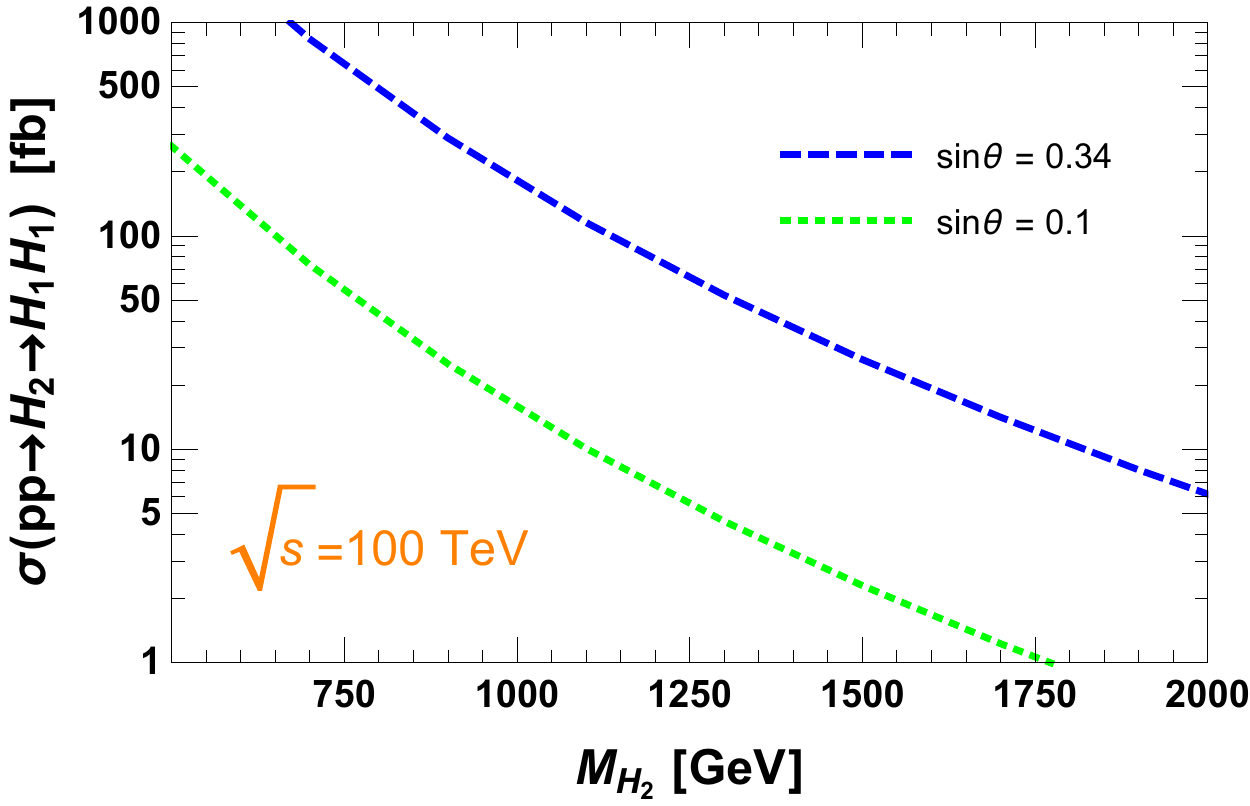}
  	\caption{Production cross section of $p p \to H_2 \to H_1 H_1$ vs mass of $H_2$ for different values of the Higgs mixing angle.} \label{fig:massvscross}	
  \end{figure}
  
  	\begin{figure}[t]
  		\centering
  		\includegraphics[width=0.75\textwidth,height=0.35\textheight]{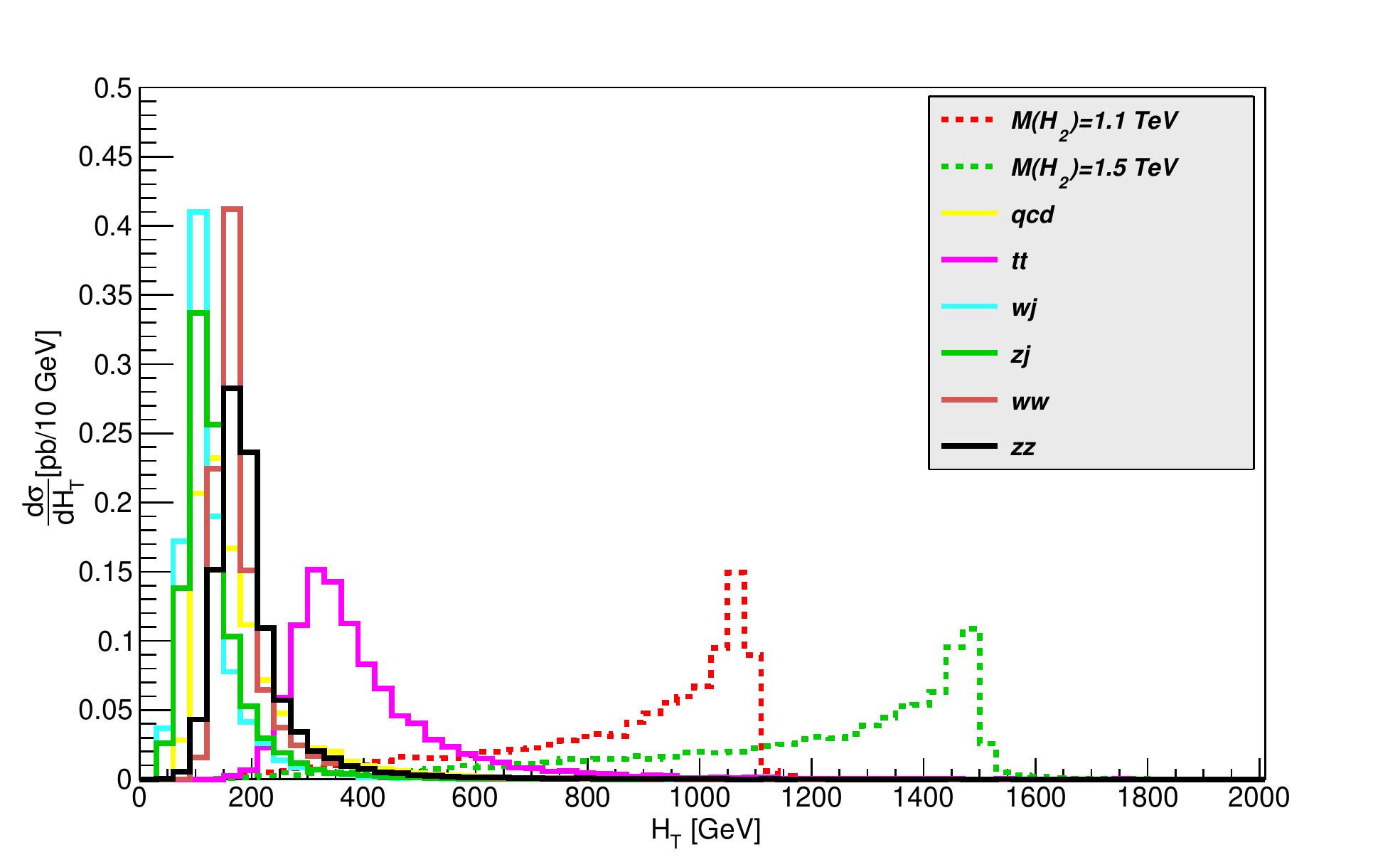}
  		\caption{$H_T$ distribution for signal and background samples.} 
  		\label{HTanddeltaeta}
  	\end{figure}
  \begin{figure}[t]
  	\centering
  	\includegraphics[width=0.45\textwidth]{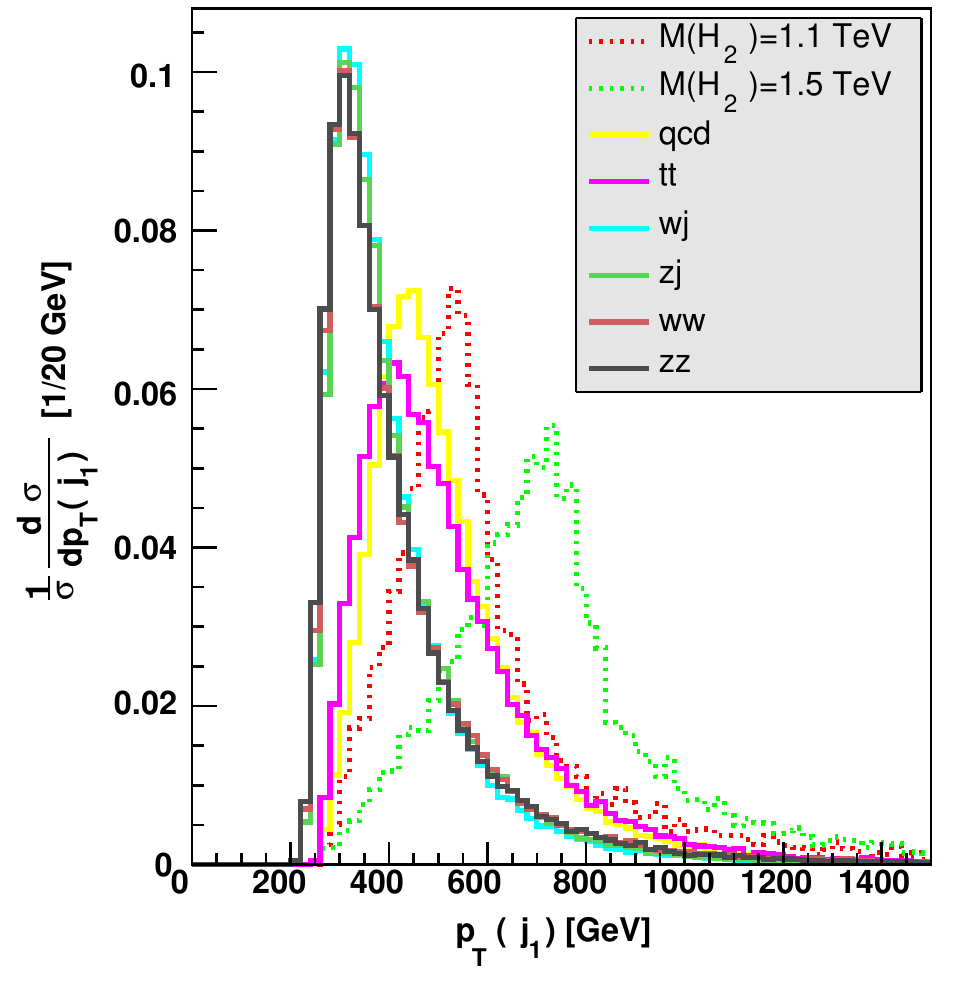} 	\includegraphics[width=0.45\textwidth]{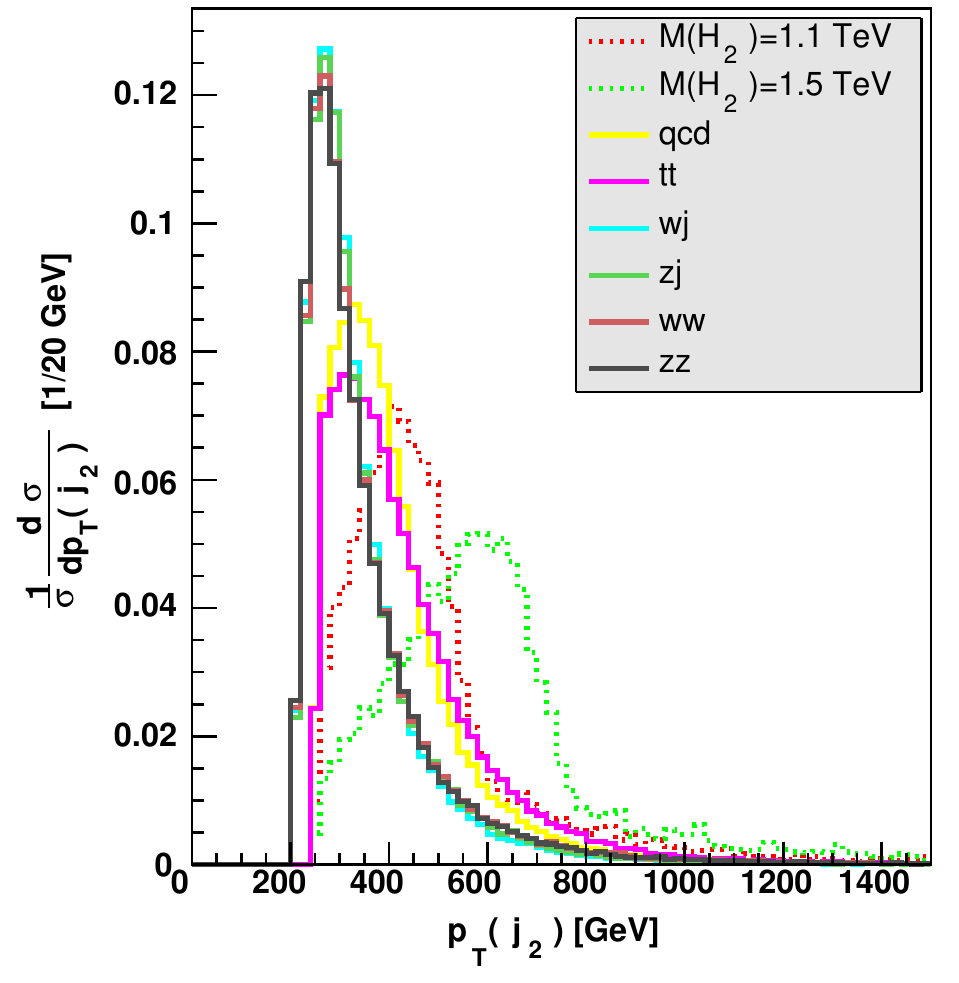} 
  	\caption{$p_T$ distribution of leading  fatjet $j_1$ (Left panel) and sub-leading fatjet $j_2$ (Right panel). We also show the $p_T$ distributions of the fatjets arising from background samples.} \label{fig:ptdis}	
  \end{figure}
  
  	\begin{figure}
	\centering
  		\includegraphics[width=0.45\textwidth]{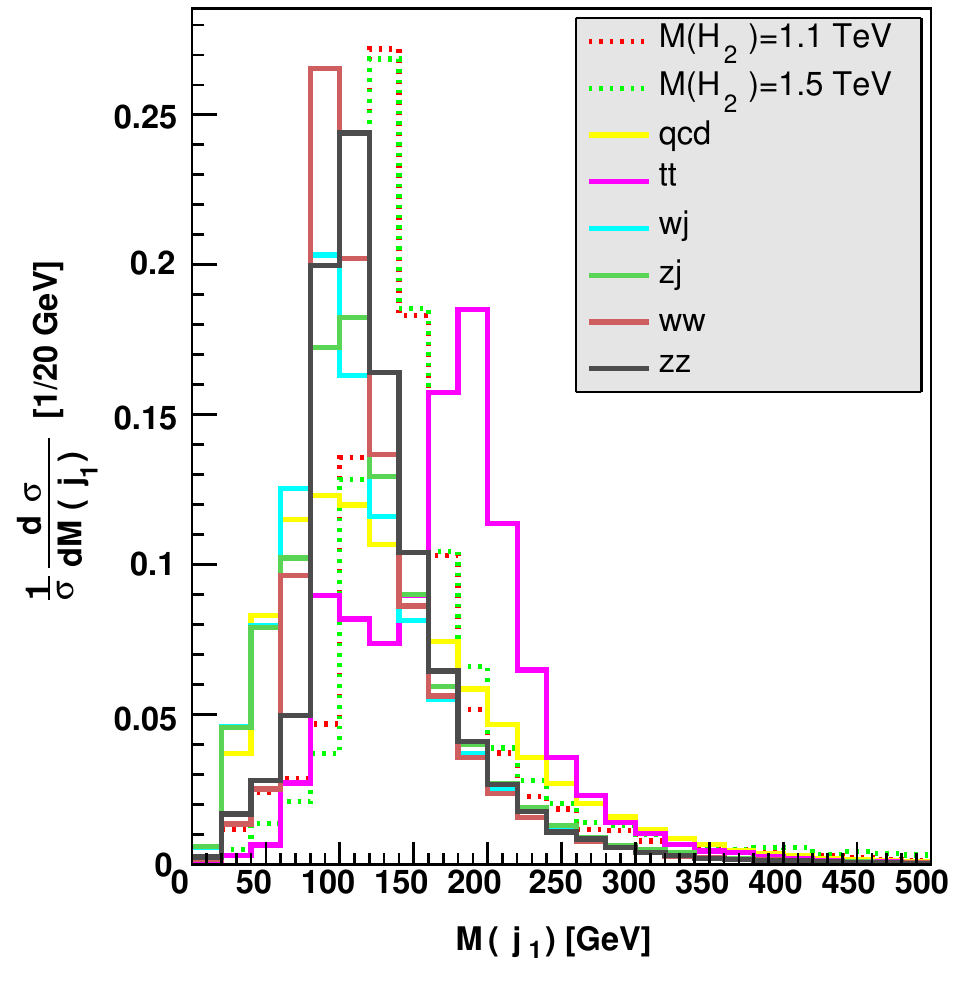}
  		\includegraphics[width=0.45\textwidth]{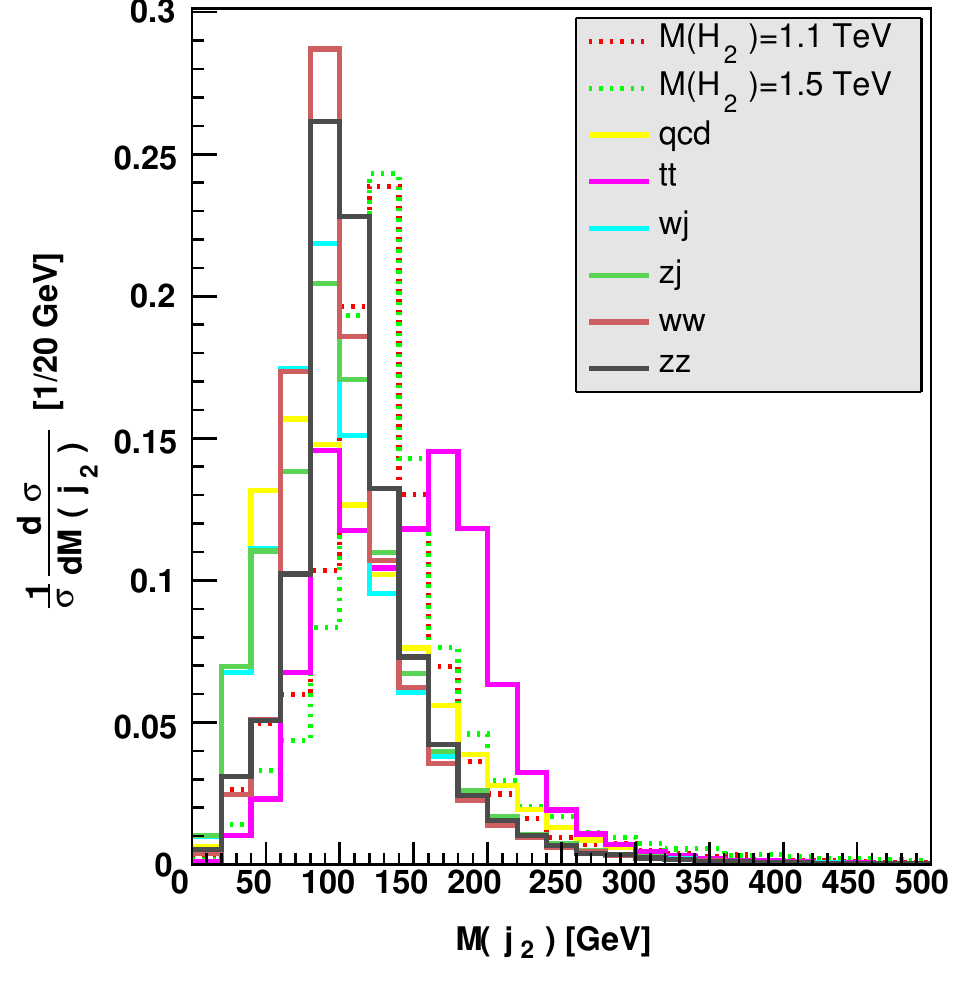}
  		\caption{Left panel: invariant mass distribution of the leading fatjet $j_1$. Right panel: the same for the sub-leading  fatjet $j_2$. } \label{fig:massdis}
  	\end{figure}

In Fig.~\ref{fig:massvscross}, we show the production cross section of $p p \to H_2 \to H_1 H_1$ at $\sqrt{s}=100$ TeV as a function of the mass of $H_2$ for $\sin\theta=0.1,\ 0.34$. There are a number of SM backgrounds, that can mimic the signal. This include both QCD and electroweak processes. The QCD is generated by combining $b\bar{b}b\bar{b}$ and $b\bar{b}jj$ final state. The other backgrounds which includes electroweak coupling are di-top ($t\bar{t}$), di-boson ($WW \text{ and } ZZ$), $Wj$ and $Zj$. Here we consider full hadronic decays of top quark, $W$ and $Z$ boson.  At the generator level, we implement these following cuts on background samples:

%\textbf{Generator level cuts:}\\
\begin{itemize}
\item The transverse momentum of the partons: $p_{T}^{j} > 20\, \textrm{GeV}$, $p_{T}^{b} > 15\, \textrm{GeV}$
\item The pseudo-rapidity of the partons: $|\eta_{j}| < 5.0 $, $|\eta_{b}| < 3.0 $
\item The separation between partons: $\Delta R_{jj} > 0.4$,  $\Delta R_{bb} > 0.2$, $\Delta R_{bj} > 0.4$
\item Invariant mass of the two $b$ quark: $m_{bb} > 30\,  \textrm{GeV}$
\item The  scalar sum of the transverse momentum of all the hadronic particles $H_T$ of the background: $H_{T} > 800\,  \textrm{GeV}$
\end{itemize}

The third cut on $\Delta R$ is to avoid any divergence, that may arise from the QCD samples. The distribution of $H_T$ is shown in Fig.~\ref{HTanddeltaeta}. The cut on $H_T$ ensures the sufficiently large background population in the desired 
region, where signal populates. We do not consider the SM di-Higgs channel into consideration, as we find that after the  $H_T$ cut, the di-Higgs channel including $h \to b \bar{b}$  branching ratio only gives $\sigma < 2.5\, \textrm{ fb}$ cross-section, which is suppressed compared to other backgrounds. The signal as compared to background shows distinct features in the distributions of different kinematic variables. In Fig.~\ref{fig:ptdis}, we show $p_T$ distributions of the two fatjets, for  two  mass points of the BSM Higgs $M_{H_2}=1.1, 1.5$ TeV.  As clearly seen in the figure 
 the $p_T$ distributions of the two leading jets for the signal and backgrounds are not very much well separated. 
 The peak of the $p_T$ distributions for the first and the second jets occurs at a relatively higher values of $p_T$ as compared to the background. Therefore, to reduce the background without affecting the
 signal we  demand a  higher values of $p_T$ on leading and sub-leading fatjets as cut,  which are $p_{T}(j_1) > 250$ GeV and $p_{T}(j_2) > 250$ GeV.   
 
\begin{figure}[t]
	\centering
  			\includegraphics[width=0.6\textwidth]{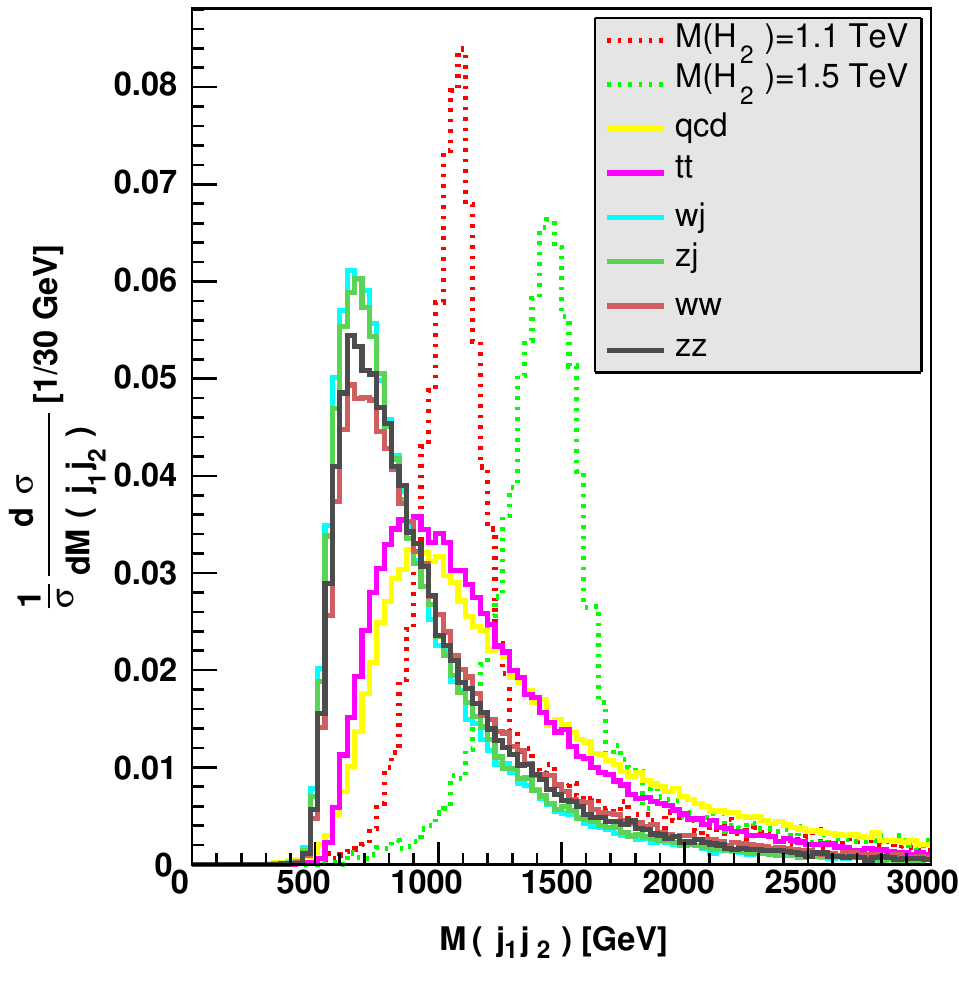}
  		\caption{Invariant mass distribution of the fatjet pair.} \label{fig:dijetmassdis}
			\end{figure}

 \paragraph{}In Fig.~\ref{fig:massdis}, the leading and sub-leading fatjet masses have been displayed. For the signal the two jets are produced from the 
 decay of the SM-like Higgs, hence the peaks for the distribution of $M(j_1)$ and $M({j_2})$ occur around the Higgs mass ($M_{H_1} \sim 125$ GeV). Since these fatjets are formed after showering and hadronization of $b\bar{b}$ pair, therefore we expect a non-trivial two prong substructure inside each of the  fatjets. We use Soft Drop algorithm~\cite{Larkoski:2014wba} which uses the condition $\frac{min(p_{T}(subjet1),p_{T}(subjet2))}{(p_{T}(subjet1)+p_{T}(subjet1))} > 0.1$ to determine whether subjets are created from Higgs decays. All subjets which satisfy this condition are qualified as the subjets originating from Higgs decay. In Delphes a subjet can not be tagged as $b$-jet. We implement a naive  $b$-jet tagging for subjets in our analysis. We use $B$-hadron to tag the subjet originating from $b$ quark. We consider $b$-tag efficiency for the subjet is $80\%$ with mis-tag efficiency $1\%$. In Fig.~\ref{fig:dijetmassdis}, we show the invariant mass distribution of the fatjet pairs, the signal peaks around $M_{H_2}$. We use these features to reduce the backgrounds while not reducing too much the signal. 
	Therefore our selection cuts are the following:
 \begin{itemize}
 	\item	$c_1$ : We  demand at least two fatjets in the final state, $N_j \geq 2$.
 	\item   $c_2$: Bound on the leading and sub-leading fatjets $p_{T}$ are 
 	$p_T(j_{1})\geq 250$ GeV \text{ and } $p_T(j_{2})\geq 250 $ GeV.
 	\item	$c_3$ : The mass of leading and sub-leading fatjet must be within 20 GeV of the SM Higgs mass, $|M_{H_1} - M_{j_{1,2}}| \leq 20$ GeV.
 	\item	$c_4$ : The invariant mass of the two fatjets will deviate at most by 150
 	GeV from the BSM Higgs mass, $|M_{H_2}-M(j_1 j_2)|\leq150$.	
 	\item $c_5$: Pseudo-rapidity separation between $j_1$ and $j_2$, $|\Delta\eta(j_1j_2)|\le 1.5$.
    \item   $c_6$ : The leading and sub-leading fatjets  must contain at least two subjets.
 	\item	$c_7$ : For the leading and sub-leading fatjets, each of the fatjets will contain two $b$-tagged subjets.
 \end{itemize}
 
We show the partonic cross-sections of different SM backgrounds in Table.~\ref{bkgSM}.  As can be seen, the main background is QCD, with a cross-section $\sigma_{QCD} \sim  10^7$ fb at the partonic level. The other backgrounds, such as $ t \bar{t}, Wj, Zj$ have a cross-section $\sigma \sim 10^4$ fb. In the third and fourth column, we show the cross-sections of the backgrounds after implementing all the cuts $c_1-c_7$.   %The third column represents the after cut cross-sections with $M(j_1 j_2)$ varied in the given range.

We also checked that  for a resolved analysis with standard set of cuts, a) number of $b$ jet $\geq 4$, b) $p_T(b) >30$ GeV, c) invariant mass of Higgs $m_{bb}= 125 \pm 20 $ GeV, d) $b$ tagging efficiency same as the fatjet analysis, and e) invariant mass of 4$b$ jet similar to the fatjet analysis, the QCD cross-section is  large $\sigma =4878$ fb (2795 fb) for $M_{H_2}=1.1$ TeV (1.5 TeV). For the resolved analysis, additional background contributions such as \textbf{$W+2j,\, Z+2j,\, jbb,$} and others are also relevant.
 \begin{table}[h]
 \centering
 	\begin{tabular}{|l|l|l|l|}
 		\hline
  BG      & $\sigma^{partonic}$ [fb]     & \thead{ $\sigma^{analysis}$ [fb]\\ ($950< M(j_1 j_2) < 1250$)}  & \thead{ $\sigma^{analysis}$ [fb]\\ ($1350< M(j_1 j_2) < 1650$)}  \\ \hline
 		 		$QCD$ &    $4.1479\times10^7$        & $1789.9$ & $211$                     \\ \hline
 		$t\bar{t}$   &    $7.603\times10^{4}$        &   $0.03$ & $9.9\times10^{-5}$                    \\ \hline
 		$Wj$ &    $5.311\times10^{4}$        & $0.0018$   & $0.43$ \\ \hline
 		$Zj$ &    $5.89\times10^{4}$        & $1.9$   & $0.006$                     \\ \hline
		$WW$ &    $ 1.2815\times10^{2}$        & $2\times10^{-6}$  & $3\times 10^{-7}$                     \\ \hline
	    $ZZ$ &    $ 3.614\times10^{1}$        & $1.3\times10^{-6}$  & $5\times10^{-8}$                     \\ \hline
 			\end{tabular}
 		\caption{Background cross section before and after the cuts.} 
		\label{bkgSM}
 \end{table}

 \begin{table}[h]
 \centering
 	\begin{tabular}{|l|l|l|l|l|}
 		\hline
 		& \multicolumn{2}{l|}{$M_{H_2}$=1.1 TeV} & \multicolumn{2}{l|}{$M_{H_2}$=1.5 TeV} \\ \hline
 		&  $\sigma^{s}$ [fb]        &      $\sigma^{b}$ [fb]     &     $\sigma^{s}$ [fb]           &  $\sigma^{b}$ [fb]           \\ \hline
 		before cut   &    $36.22 \ (3.13)$        &   $4.17\times10^7$         &  $ 8.64 \ (0.75) $          &    $4.17\times10^7$             \\ \hline
 		after cut &    $0.745 \ (0.064)$        & $1791.9$           &    $0.19 \ (0.016)$     &  $211.43$           \\ \hline
 		$\frac{\sigma^{s} \sqrt{\mathcal{L}}}{\sqrt{\sigma^{s}+\sigma^{b}}}$, $\mathcal{L}=30  \ \text{ab}^{-1}$& \multicolumn{2}{l|}{$\qquad3.05  \ (0.26)\qquad$} & \multicolumn{2}{l|}{$\qquad2.26 \ (0.19)\qquad$} \\ \hline
 			\end{tabular}
 		\caption{Signal ($pp \to 2 j_\textrm{fat}$) and background cross-sections after different selection cuts at $\sqrt{s}=100$ TeV for $\sin\theta=0.34 \ (0.1)$. } 
 		\label{cba-table}
 \end{table}
% \end{center}
 In Table \ref{cba-table}, we  show the signal  cross-sections  before and after applying the cuts $c_{1}-c_7$. We consider  two illustrative BSM Higgs masses $M_{H_2}=1.1,1.5$ TeV.  The background in this table corresponds to the QCD background, shown in Table.~\ref{bkgSM}, as this is the major background.  As can be seen from the table the background is huge as compared to the signal. Applying the cuts however allows to improve the significance of the
 signal. The main  remaining  background is $QCD$. We find  that for $M_{H_2}$=1.1 TeV , one can achieve
 $3\sigma$ significance of the signal over background for 30 $ab^{-1}$ luminosity. For higher mass values of the
 BSM Higgs, significance reduces.   The significance of the signal can be improved over the background if one uses multivariate analysis and neural network methods. Additional final states such as $b\bar{b}\gamma \gamma$/ $b\bar{b}\tau^+ \tau^-$ are expected to give better significance, as these are clean channels. Detailed evaluation of the discovery prospect of all these channels is beyond the scope of this paper, and we will present this elsewhere.
  
\section{Conclusion \label{conclu}}
In this work, we adopt  an effective field theory framework that contains RHN and one  SM gauge singlet  scalar. Our model accommodates  a FIMP DM candidate and explains the observed eV  masses of light neutrinos, where SM neutrinos acquire their masses via seesaw mechanism. Three gauge singlet RHN states $N_{1,2,3}$ and one  gauge singlet real scalar $\chi$ are present in addition to the SM fields. The two RHN states  $N_{1,2}$ participate  in light neutrino mass generation and $N_3$ is the DM. There is sizeable mixing between the BSM scalar and the SM Higgs, that offers better detection prospect of the BSM Higgs at colliders. 
\paragraph{}The FIMP DM candidate $N_3$ in our model interacts with the SM and BSM scalars only via effective $d=5$ Yukawa interaction. The $d=5$ term generates the tri-linear interaction term responsible for decay once $\chi$ and $\Phi$ acquire \textit{vev}. Hence FIMP DM can be produced from the decay of the SM and BSM Higgs. Annihilation of scalars and other SM particles can also lead to DM production. However,  for a low reheating temperature, the decay contribution dominates. In our analysis, we therefore first consider a low reheating temperature and analyse only decay contributions. In {\it Scenario-I and II}, where there is no bare-mass term of $N_{1,2,3}$ being added, both the DM mass and its interaction with other particles depend on the same operator. Therefore, the relic density constraint leads to a  strong correlation between the \textit{vev} of BSM scalar ($v_\chi$) and mass of DM ($M_{N_{3}}$). Keeping other parameters fixed, the required value of  $v_\chi$ to satisfy the observed relic density increases with the mass of DM. The same $v_\chi$ also primarily governs the BSM Higgs mass $M_{H_2}$. Since in our model SM and BSM Higgs mixing can be sizeable, a TeV scale or lighter $H_2$ therefore has better discovery prospect at collider as compared to a very heavy $H_2$. We find that, for TeV scale $v_{\chi}$ which is a natural choice for TeV scale or lower BSM Higgs state,  DM relic density constraint is satisfied only if its mass is in the KeV range. We also consider another scenario {\it  Scenario-III}, where we accommodate a bare mass term of the RHN states. We find that in this case, the  tight correlation between {\it vev} of $\chi$ and mass of DM is somewhat relaxed, and a GeV scale DM is possible to accommodate with a TeV scale BSM Higgs/{\it vev} $v_{\chi}$. 
\paragraph{}We also consider a variation of the reheating temperature $T_R$ and study the different annihilation channels. For a high reheating temperature we consider both the decay and annihilation contributions in relic density, where the latter dominates the relic abundance. A number of annihilation channels $WW/ZZ/H_iH_i \to N_3 N_3$ can give significant contributions. In our analysis, we show the variation of relic density w.r.t various parameters, such as $v_{\chi}$, mass of DM, and  reheating temperature. We find that the relic density  increases with the mass of DM, and $T_R$ (for gauge boson and scalar annihilation only), and decreases for higher {\it vev} of BSM scalar. Assuming BSM Higgs varying in $\mathcal{O}(\textrm{TeV})$ range, we vary these parameters in a wide range and show the variation of relic density as scatter plot. 
\paragraph{}Finally, we explore the collider signature of the TeV scale BSM scalar  at the 100 TeV future $pp$ machine. We consider the production of  the  BSM scalar which has sizeable mixing with the SM Higgs, and its decay to a pair of SM Higgs states. We further consider the decay of the SM Higgs to $b\bar{b}$ states. For a TeV scale heavy Higgs, the SM Higgs is rather moderately boosted leading to collimated decay products. We consider di-fatjet final states as our model signature. We perform a detailed analysis considering several backgrounds, such as, QCD, $t \bar{t}$, $WW/ZZ, W+1j, Z+1j$. Following a cut based analysis we find that a $3\sigma$ significance can be achieved for a 1.1 TeV BSM scalar with $30\text{ ab}^{-1}$ luminosity. Thus,  the di-fatjet channel which is sensitive to the tri-linear Higgs coupling $H_2H_1H_1$ is a complementary probe for the heavy BSM Higgs, in addition to other channels, such as $pp\to H_2\to WW/ZZ$.
\section*{Acknowledgments}
%%%%%%%%%%%%%%%%%%%%%%%%%%%%%%%%%%%%%%%%%%%%%%%%%%%%%%%%%%%%%%%%%%%%%%%%%
G.B and M.M acknowledge  the support from the Indo-French Centre for the Promotion of Advanced Research (Grant no: 6304-2). M.M thanks DST INSPIRE Faculty research grant (IFA-14-PH-99). S.K would like to thank cluster computing facility at GWDG, G\"{o}ttingen. S.S and R.P acknowledge  the support of the SAMKHYA: High Performance Computing Facility provided by IOPB. M.M, R.P and S.S thank Dr. Shankha Banerjee for useful discussions on  di-Higgs searches.  
\section*{Appendix \label{App:AppendixA}}
In this section, we discuss various expressions of the annihilation contributions to relic density. 
\subsection*{Analytical expression of relevant cross sections}
We list the cross-sections  for the $A B \rightarrow N_{3} N_{3}$ processes where $A$, $B$ are any SM particles contributing to DM production in the freeze-in mechanism.
\begin{itemize}
\item {\bf $WW \rightarrow N_{3} N_{3}$: }
\begin{eqnarray}
g_{H_1 W^{+} W^{-}} &=& \frac{M_W e \cos \theta}{s_w}\,,\nn \\
g_{H_2 W^{+} W^{-}} &=& \frac{M_W e \sin \theta}{s_w}, \nn \\
A_{WW} &=& \dfrac{g_{H_1 W^{+} W^{-}}\,\,g_{H_{1}N_{3} N_{3}}}{(s-M_{H_1}^{2}) + i M_{H_1} \Gamma_{H_1}}
+ \dfrac{g_{H_2 W^{+} W^{-}}\,g_{H_{2} N_{3} N_{3}}}
{(s-M_{H_2}^{2}) + i M_{H_2} \Gamma_{H_2}}, \nn \\
M_{WW} &=& \dfrac{2}{9}\,\left(1 + \dfrac{(s - 2M_{W}^{2})^{2}}
{8M_{W}^{4}}\right) \,\, \left( s - 4 M^2_{N_3} \right)\,|A_{WW}|^{2}, \nn \\
\sigma_{WW \rightarrow N_{3} N_{3}} &=& \dfrac{1}{16 \pi s}\,\,
\sqrt{\dfrac{s - 4M_{N_3}^{2}}{s - 4M_{W}^{2}}} \,\,\,M_{WW}\,.
\end{eqnarray}

\item {\bf $ZZ \rightarrow N_{3} N_{3}$: }
\begin{eqnarray}
g_{H_1 ZZ} &=& \frac{M_W e \cos \theta}{c^2_w s_w}\,,\nn \\
g_{H_2 ZZ} &=& \frac{M_W e \sin \theta}{c^2_w s_w}, \nn \\
A_{ZZ} &=& \dfrac{g_{H_1 ZZ}\,\,g_{H_{1}N_{3} N_{3}}}{(s-M_{H_1}^{2}) + i M_{H_1} \Gamma_{H_1}}
+ \dfrac{g_{H_2 ZZ}\,g_{H_{2} N_{3} N_{3}}}
{(s-M_{H_2}^{2}) + i M_{H_2} \Gamma_{H_2}}, \nn \\
M_{ZZ} &=& \dfrac{2}{9}\,\left(1 + \dfrac{(s - 2M_{W}^{2})^{2}}
{8M_{W}^{4}}\right) \,\, \left( s - 4 M^2_{N_3} \right)\,|A_{ZZ}|^{2}, \nn \\
\sigma_{ZZ \rightarrow N_{3} N_{3}} &=& \dfrac{1}{32 \pi s}\,\,
\sqrt{\dfrac{s - 4M_{N_3}^{2}}{s - 4M_{Z}^{2}}} \,\,\,M_{ZZ}\,.
\end{eqnarray}

\item {\bf $f \bar{f} \rightarrow N_{3} N_{3}$: }
\begin{eqnarray}
g_{H_1 ff} &=& - \frac{e M_f \cos \theta}{2 M_W}\,,\nn \\
g_{H_2 ff} &=& - \frac{e M_f \sin \theta}{2 M_W}, \nn \\
A_{ff} &=& \dfrac{g_{H_1 ff}\,\,g_{H_{1}N_{3} N_{3}}}{(s-M_{H_1}^{2}) + i M_{H_1} \Gamma_{H_1}}
+ \dfrac{g_{H_2 ff}\,g_{H_{2} N_{3} N_{3}}}
{(s-M_{H_2}^{2}) + i M_{H_2} \Gamma_{H_2}}, \nn \\
M_{ff} &=& \dfrac{2}{n_c}\,\left( s - 4 M^2_{f}\right) \,\, \left( s - 4 M^2_{N_3} \right)\,|A_{ff}|^{2}, \nn \\
\sigma_{f \bar{f} \rightarrow N_{3} N_{3}} &=& \dfrac{1}{64 \pi s}\,\,
\sqrt{\dfrac{s - 4M_{N_3}^{2}}{s - 4M_{f}^{2}}} \,\,\,M_{ff}\,.
\end{eqnarray}
\item {\bf $H_{i} H_{j} \rightarrow N_{3} N_{3}\,\, (i,j = 1, 2)$}:

\begin{eqnarray}
&&g_{H_{1}H_{1}H_{1}} = -3\,[2\,v\lambda_{H_1}\cos^{3}\theta
+ 2\,v_{\chi}\,\lambda_{H_2}\sin^{3}\theta +
\lambda_{H_{1}H_{2}}\sin\theta\,\cos\theta\,
(v\sin\theta + v_{\chi}\cos\theta)],\nn \\
&&g_{H_{1}H_{1}H_{2}} = [6\,v\lambda_{H_1}\cos^{2}\theta\sin\theta -
6\,v_{\chi}\lambda_{H_2}\sin^{2}\theta\,\cos\theta 
-(2-3\,\sin^{2}\theta)\,v\,\lambda_{H_{1}H_{2}}\,\sin\theta
\nn\\ 
&&~~~~~~~~~-(1-3\sin^{2}\theta)v_{\chi}\,
\lambda_{H_{1}H_{2}}\cos\theta] \,,\label{h2h1h1} \\
&&g_{H_{2}H_{2}H_{2}} = 3\,[2\,v\lambda_{H_1}\sin^{3}
\theta - 2\,v_{\chi}\lambda_{H_2}\cos^{3}\theta +
\lambda_{H_{1}H_{2}}\sin\theta\cos\theta\,
(v\cos\theta - v_{\chi}\sin\theta)],\nn \\
&&g_{H_{2}H_{2}H_{1}} = -[6\,v\lambda_{H_1}\sin^{2}
\theta\cos\theta + 6\,v_{\chi}\lambda_{H_2}\cos^{2}\theta\sin\theta 
-(2-3\,\sin^{2}\theta)v_{\chi}\lambda_{H_{1}H_{2}}\sin\theta
\,\,\nn \\
&&~~~~~~~~~+(1-3\sin^{2}\theta)v\lambda_{H_{1}H_{2}}\cos\theta]\,, \nn \\
&&g_{H_{1}H_{1} N_{3} N_{3}} =
(\frac{c^{\prime}_{33}}{\Lambda} \cos^{2}\theta + \frac{c_{33}}{\Lambda} \sin^{2}) \,, \nn \\
&&g_{H_{2}H_{2} N_{3} N_{3}} = (\frac{c^{\prime}_{33}}{\Lambda} \sin^{2}\theta + \frac{c_{33}}{\Lambda} \cos^{2}\theta) \nn \\
&&g_{H_{1}H_{2}N_{3} N_{3}} = \sin\alpha \cos\alpha(\frac{c^{\prime}_{33}}{\Lambda} -
\frac{c_{33}}{\Lambda}) \,,\nn \\
%\end{eqnarray}
%\begin{eqnarray}
&&M_{H_{i} H_{j}} = \dfrac{g_{H_{i}H_{j}H_{1}}\,
\,g_{H_{1} N_{3} N_{3}}}{(s-M_{H_1}^{2}) +
i M_{H_1} \Gamma_{H_1}} + \dfrac{g_{H_{i}H_{j}H_{2}}
\,\,g_{H_{2} N_{3} N_{3}}}{(s-M_{H_2}^{2}) +
i M_{H_2} \Gamma_{H_2}} + g_{H_{i}H_{j} N_{3} N_{3}}, \nn \\
&&\sigma_{H_{i}H_{j} \rightarrow N_{3} N_{3}} =
\dfrac{1}{16 \pi s}\,\,\sqrt{\dfrac{s(s - 4M_{N_3}^{2})}
{(s-(M_{H_i}+M_{H_j})^{2})(s - (M_{H_i} - M_{H_j})^{2})}}
\,\,\,|M_{H_{i}H_{j}}|^{2}\,.
\end{eqnarray}

\end{itemize}

As given in \cite{Hall:2009bx}, we can approximately calculate the DM contribution analytically for higher value of the reheating
temperature by solving the following Boltzmann equation,
\begin{eqnarray}
\frac{d n_{N_3}}{d t} + 3 n_{N_3} H \simeq \frac{T}{1024} \int ds d\Omega \sqrt{s} |M|^{2}_{AB \rightarrow N_{3} N_{3}} 
K_{1}\left(\frac{\sqrt{s}}{T} \right)\,.
\end{eqnarray}
We parametrise the amplitude for the $A\,B \rightarrow N_{3} N_{3}$ process to be proportional to some power of
centre of mass energy at very high temperature {\it i.e.} $|M|^{2}_{AB \rightarrow N_{3} N_{3}} = \alpha s^{n}$ where 
$\alpha$ is a constant which depends on the couplings and $n$ is a rational number. After substituting the above amplitude and focusing on the dependence of the co-moving number density on temperature, we obtain
\begin{eqnarray}
\frac{d Y_{UV}}{d T} = \kappa T^{2 n - 2}
\end{eqnarray}
where $Y_{UV}$ is the co-moving number density of DM and $\kappa$ is a constant.
In the present work the amplitude varies in the following way at large $s$,
\begin{eqnarray}
|M|^{2} &=& \alpha_{AB} s \,\,\,{\rm where}\,\,\,A, B\,\, =\,\,W, Z, H_1, H_2 \nn \\
 &=& \alpha_{ff} \,\,\,{\rm where}\,\,\,{f \,\, is\,\, \rm SM\,\, \rm fermion}\,. 
\end{eqnarray}  

Therefore, for gauge bosons and Higgses we can easily show that,
\begin{eqnarray}
Y_{UV} = \kappa T_{R}
\label{dependence-yuv-TR}
\end{eqnarray}
and for the fermion it will be
\begin{eqnarray}
Y_{UV} = \kappa( \frac{1}{T_0} - \frac{1}{T_R})\,.
\label{dependence-ff-yuv-TR}
\end{eqnarray}

Finally, we can conclude that for the Higgs bosons and gauge boson the relic density contribution increase linearly with $T_R$, whereas for fermions the contribution does not change significantly with the $T_R$ for high value of $T_R$.
\bibliographystyle{utphys}
\bibliography{bibitem}  	
\end{document}